\documentclass[11pt]{article}
\pdfoutput=1

\usepackage{ae}
\usepackage[T1]{fontenc}
\usepackage[ansinew]{inputenc}
\usepackage{mathrsfs}
\usepackage{amsmath}
\numberwithin{equation}{section}
\usepackage{amssymb}
\usepackage{dsfont}
\usepackage{graphicx}
\usepackage{upgreek}
\usepackage{color}
\usepackage{pst-all}
\usepackage{tikz}
\usepackage{import}
\usepackage{hyperref}
\usepackage{cite}

\usepackage{epsfig}
\usepackage{wasysym}

\usepackage{verbatim}
\usepackage{stmaryrd}

\usepackage{amsfonts}
\usepackage{xcolor}

\usepackage{tikz}
\usetikzlibrary{arrows,shapes}
\usetikzlibrary{patterns,fadings}

\usetikzlibrary{snakes,arrows,shapes,positioning,decorations}

\usepackage[a4paper, total={6.5in, 9.3in}]{geometry}
\usepackage{dsfont}
\usepackage{empheq}
\usepackage{slashed}
\usepackage{subcaption}
\usepackage{multirow}
\allowdisplaybreaks

\tikzset{
BPSbox/.style={
       fill={rgb,255: red,221; green,221; blue,221},
       draw={rgb,255: red,113; green,113; blue,113}, thick,
       text width=2.2cm,
       inner sep=2pt,
       text centered,
       }}
\tikzset{ BPSbox2/.style={
       fill={rgb,255: red,221; green,221; blue,221},
       draw={rgb,255: red,113; green,113; blue,113}, thick,
       text width=2.68cm,
       inner sep=2pt,
       text centered,
       }
}

\begin{document}

\thispagestyle{empty}

\renewcommand{\thefootnote}{\fnsymbol{footnote}}
\setcounter{footnote}{0}
\setcounter{figure}{0}
\begin{flushright}
{MPP-2023-41}
\end{flushright}
\vskip 0.3in
\begin{center}

{\Large
\textbf{\mathversion{bold}
Integrated negative geometries in ABJM
}}

\vspace{1.0cm}

\textrm{Johannes M. Henn$^{a,}$\footnote{henn@mpp.mpg.de}, Mart\'in Lagares$^{a,b,}$\footnote{martinlagares95@gmail.com} and Shun-Qing Zhang$^{a,}$\footnote{sqzhang@mpp.mpg.de}}
\\ \vspace{1.2cm}
\footnotesize{

\textit{$^a$ Max-Planck-Institut f\"ur Physik, Werner-Heisenberg-Institut, 80805 M\"unchen, Germany.}

\textit{$^b$Instituto de F\'isica La Plata, Argentina} }

\vspace{1cm}

\vspace{3mm}

\par\vspace{1.5cm}

\textbf{Abstract}\vspace{2mm}
\end{center}

We study, in the context of the three-dimensional ${\cal N}=6$ Chern-Simons-matter  (ABJM) theory, the infrared-finite functions that result from performing $L-1$ loop integrations over the $L$-loop integrand of the logarithm of the four-particle scattering amplitude. Our starting point are the integrands obtained from the recently proposed all-loop projected amplituhedron for the ABJM theory. 
Organizing them in terms of negative geometries ensures that no divergences occur upon integration if at least one loop variable is left unintegrated.
We explicitly perform the integrations up to $L=3$, finding both parity-even and -odd terms. Moreover, we discuss a prescription to compute the cusp anomalous dimension $\Gamma_{\rm cusp}$ of ABJM in terms of the integrated negative geometries, and we use it to reproduce the first non-trivial order of $\Gamma_{\rm cusp}$. Finally, we show that the leading singularities that characterize the integrated results are conformally invariant.

\noindent

\vspace*{\fill}

\setcounter{page}{1}
\renewcommand{\thefootnote}{\arabic{footnote}}
\setcounter{footnote}{0}

\newpage
\tableofcontents

%%%%%%%%%%%%%%%%%%%%%%%%%%%%%%%%%%%%%%%%%%%%%%%%%%%%%%%%%%%%%%%%%%%%%%%%%%%%%%%%%%%%%%%%%%%%%%%%%%%%%%%%%%%%%%%%%%%%%%%%%%%%%%%%%%%%%%%%%%%%%%%%%%%%%%%%%%%%%%%%%%%%%%%%%%%%%%
\section{Introduction}

Over the last decades collider physics has pushed the threshold of precision in experimental data for particle physics to unprecedented values. Naturally, these results have stimulated the study of scattering amplitudes, leading to substantial developments in the field over the last years. In particular, lots of efforts have been put in trying to explain the surprising simplicity that those observables often show, with the Parke-Taylor formula\cite{Parke:1986gb} being the first of many examples. Computations with hundreds of Feynman diagrams often lead to results that can be written within a line, suggesting an underlying simplicity that should be understood.
A case in point is the discovery of positive geometries \cite{Arkani-Hamed:2013jha,Arkani-Hamed:2013kca,Franco:2014csa,Arkani-Hamed:2017vfh,Arkani-Hamed:2018rsk,Damgaard:2019ztj,Ferro:2022abq}, that provide a geometrical interpretation for scattering amplitudes in numerous quantum field theories.

In this new geometric picture, the $S$-matrix of the four-dimensional $\mathcal{N}=4$ super Yang-Mills (sYM) theory is viewed, in the planar limit, as the \textit{volume} of a mathematical object whose boundaries encode the physical singularities of the amplitudes. More precisely, it has been shown that the canonical form of a certain positive geometry, the so called \textit{Amplituhedron}, gives the tree-level amplitudes and all-loop integrands for an arbitrary $n$-particle scattering process in the planar $\mathcal{N}=4$ sYM theory \cite{Arkani-Hamed:2013jha}. Interestingly, in this new framework concepts such as locality and unitarity are no longer fundamental principles to be assumed, but rather they are derived properties. This geometric formulation of scattering amplitudes has also been applied in other contexts, such as the bi-adjoint $\phi^3$ theory \cite{Arkani-Hamed:2017mur}, cosmology \cite{Arkani-Hamed:2017fdk} or, as it will be of interest for us, the three-dimensional ${\cal N}=6$ Chern-Simons-matter theory known as ABJM\cite{Huang:2021jlh,He:2021llb,He:2022cup}. For recent reviews about the study of amplitudes in terms of positive geometries see \cite{Ferro:2020ygk,Herrmann:2022nkh}.

Recently, the authors of \cite{Arkani-Hamed:2021iya} proposed a novel way to express the ${\cal N}=4$ sYM amplituhedron as a sum over \textit{negative} geometries. The latter, characterized by a change of sign in the defining inequalities, naturally give rise to the logarithm of the amplitude: its integrand at each loop order is simply given by summing over a certain subset of negative geometries, represented by \textit{connected graphs} 
in the pictorial representation of \cite{Arkani-Hamed:2021iya}. 

As is well known, infrared divergences exponentiate in planar Yang-Mills theories. In particular, in conformal field theories the logarithm of an amplitude has only double poles $1/\epsilon^2$ in the dimensional regulator $D=4-2\epsilon$ (with the coefficient being the cusp anomalous dimension), while the $L$-loop amplitude has poles of order $1/\epsilon^{2L}$.
In the ${\cal N}=4$ sYM theory there is a naturally related quantity that is completely free of divergences.
It arises when one considers the integration of the $L$-loop negative geometry over $L-1$ of the loop variables, i.e. leaving one of the loop variables unintegrated \cite{Arkani-Hamed:2021iya,Alday:2011ga,Adamo:2011cd,Engelund:2011fg,Alday:2012hy,Alday:2013ip,Hernandez:2013kb,Henn:2019swt,Chicherin:2022bov,Chicherin:2022zxo}. Remarkably, this object is infrared (IR) finite, and all the divergences concentrate on the $L$-th loop integral. This can be seen as a consequence of organizing the results as a sum of negative geometries \cite{Arkani-Hamed:2021iya}. Moreover, one can show that the result of integrating $L-1$ of the loop variables can be expressed in terms of a function of $3n-11$ conformal cross-ratios. This is the same number of kinematic variables as for QCD $n$-point amplitudes. This similarity, together with a conjectured duality with pure Yang-Mills all-plus helicity-amplitudes \cite{Chicherin:2022bov}, motivates further studies of these finite observables. We will focus on the four-particle case, for which one gets a function $\mathcal{F}(z)$ of a single cross-ratio.

An exciting outcome of the study of $\mathcal{F}(z)$ comes when taking into account the duality between scattering amplitudes and Wilson loops in ${\cal N}=4$ sYM \cite{Alday:2007hr,Drummond:2007aua,Brandhuber:2007yx,Drummond:2007cf}. Interestingly, this duality allows to recover the $L$-loop contribution to the cusp anomalous dimension $\Gamma_{\rm cusp}$ from the $(L-1)$-loop term in the perturbative expansion of $\mathcal{F}(z)$ \cite{Arkani-Hamed:2021iya,Alday:2013ip,Henn:2019swt}. This prescription has been used to compute the full four-loop contribution to $\Gamma_{\rm cusp}$ both in ${\cal N}=4$ sYM and in QCD, including the first non-planar corrections \cite{Henn:2019swt}.

Besides of the fact that it is IR-finite, many other interesting properties and results have been found for $\mathcal{F}(z)$. As shown by \cite{Arkani-Hamed:2021iya}, in ${\cal N}=4$ sYM one can perform a non-perturbative sum over a particular subset of negative geometries (more precisely, ladder- and tree-type diagrams), opening the door for a full all-loop computation of $\mathcal{F}(z)$. Such results would allow a comparison with the non-perturbative derivation of $\Gamma_{\rm cusp}$ coming from integrability \cite{Beisert:2006ez,Correa:2012hh}. Also surprisingly, the leading singularities of these integrated negative geometries enjoy a (hidden) conformal symmetry \cite{Chicherin:2022bov,Chicherin:2022zxo}. 
Furthermore, identities relating $\mathcal{F}(z)$ to all-plus amplitudes in pure Yang-Mills theory have been found \cite{Chicherin:2022bov,Chicherin:2022zxo}. Finally, one can also note that the perturbative expansion of $\mathcal{F}(z)$ respects a uniform transcendentality principle \cite{Henn:2019swt}.

Taking into account the previous considerations, it seems natural to pose the question of how the above results generalize to other theories, ultimately aiming for a generalization to QCD. In this regard, the three-dimensional ABJM theory \cite{Aharony:2008ug} emerges as a reasonable candidate, given its well-known similarities with ${\cal N}=4$ sYM. Much progress has been made in understanding the properties of scattering amplitudes in this three-dimensional case. The four-particle scattering amplitude is known up to three-loops \cite{Bargheer:2010hn,Chen:2011vv,Bargheer:2012cp,Leoni:2015zxa,Bianchi:2014iia}, and there is a BDS-like conjecture for the all-loop result \cite{Bianchi:2011aa,Bianchi:2014iia}. Moreover, even non-planar corrections have been computed \cite{Bianchi:2013iha,Bianchi:2013pfa}. For $n=6$ and $n=8$ particles the current frontier is two-loops \cite{Caron-Huot:2012sos,He:2022lfz}, and there are one-loop results for scattering processes with arbitrary number of particles \cite{Bianchi:2012cq}. Furthermore, the Wilson loops/scattering amplitudes duality is believed to hold for the four-particle case, but has been shown to fail when the number of particles increases \cite{Henn:2010ps,Bianchi:2011dg,Chen:2011vv,Leoni:2015zxa}. There is also evidence of dual-superconformal \cite{Huang:2010qy,Gang:2010gy,Chen:2011vv,Bianchi:2011fc} and Yangian symmetry\cite{Bargheer:2010hn,Lee:2010du}.

The geometric formulation of amplitudes in terms of positive geometries was first extended at tree level to the ABJM theory in \cite{Huang:2021jlh,He:2021llb}. Recently, the authors of \cite{He:2022cup} proposed an all-loop \textit{projected amplituhedron} for ABJM by imposing a symplectic condition on the amplituhedron of ${\cal N}=4$ sYM. This conjecture has been checked up to $L=5$ loops for the four-particle case. Along the lines of \cite{Arkani-Hamed:2021iya}, the projected amplituhedron allows for a decomposition in terms of negative geometries. More importantly, comparing to the four-dimensional case, in three dimensions a smaller number of negative geometries contribute to the integrand of the logarithm of the amplitude. More precisely, only those geometries associated to \textit{bipartite} graphs contribute, allowing for a significant simplification in the perturbative expansion of the integrand. We should note that the study of integrated negative geometries in ABJM is interesting towards an all-loop computation of the ABJM cusp anomalous dimension\cite{Gromov:2008qe,Griguolo:2012iq,Bianchi:2014ada,Bianchi:2013pfa}. Non-perturbative results would clear the way for the all-loop computation of the interpolating function $h(\lambda)$ of ABJM \cite{Beisert:2006qh,Minahan:2008hf,Minahan:2009te,Bak:2008vd,Minahan:2009aq,Minahan:2009wg,Leoni:2010tb,Gromov:2014eha,Cavaglia:2016ide}, whose knowledge is crucial to exploit the results coming from integrability. An all-loop expression for $h(\lambda)$ was proposed in \cite{Gromov:2014eha,Cavaglia:2016ide}. 

In this paper we focus on the ABJM theory, and we explicitly perform the $(L-1)$-loop integrations of the four-particle negative geometries for the $L \leq 3$ cases, showing that the integrated results are given by finite and uniform-transcendental polylogarithmic functions. In an analogous way to the five-particle case of ${\cal N}=4$ sYM \cite{Chicherin:2022zxo}, we find it convenient to organize the integrated results in parity-even and -odd terms,  which are described by two functions $\mathcal{F}(z)$ and $\mathcal{G}(z)$, respectively.  As we will see, it is straightforward to show that only the former contributes to the cusp anomalous dimension $\Gamma_{\rm cusp}$ after the last loop integration. Furthermore, we use our results to compute the first non-trivial contribution to $\Gamma_{\rm cusp}$, finding perfect agreement with the literature \cite{Griguolo:2012iq}. Finally, we discover that the leading singularities of the integrated results also possess a hidden conformal symmetry, in a similar manner to what was found in the four-dimensional case \cite{Chicherin:2022bov,Chicherin:2022zxo}. 

The paper is organized as follows. In Section \ref{sec: Integrands from negative geometries} we review the role that negative geometries play in the construction of integrands in ${\cal N}=4$ sYM and in ABJM, and we finish with a discussion of how dual conformal invariance constrains the expressions that come from integrating these geometries. Then, in Section \ref{sec: perturvative results} we perform, up to two loops, the explicit integration of the negative geometries of ABJM. In Section \ref{sec: cusp anomalous dimension} we discuss how one can compute the cusp anomalous dimension of ABJM as a consequence of applying a functional on the integrated negative geometries. In Section \ref{sec:transcendental} we turn to the analysis of the transcendental weight properties of our results. Section \ref{sec: Conformal invariance of leading singularities} is devoted to the symmetry analysis of the leading singularities that characterize the integrated results. We give our conclusions in Section \ref{sec: conclusions}. Finally, there are three appendices that complement the results discussed in the main body of the paper.

%%%%%%%%%%%%%%%%%%%%%%%%%%%%%%%%%%%%%%%%%%%%%%%%%%%%%%%%%%%%%%%%%%%%%%%%%%%%%%%%%%%%%%%%%%%%%%%%%%%%%%%%%%%%%%%%%%%%%%%%%%%%%%%%%%%%%%%%%%%%%%%%%%%%%%%%%%%%%%%%%%%%%%%%%%%%%%
\section{Integrands from negative geometries}
\label{sec: Integrands from negative geometries}

In this paper we will analyze, within the context of the ABJM theory, the behaviour of $L$-loop integrands for the logarithm of the amplitude after performing $L-1$ of the corresponding loop integrations. Therefore, it is instructive to review how one can express the aforementioned integrands in terms of canonical forms of negative geometries.

Let us consider a $D$-dimensional gauge theory, with focus on a four-particle scattering process of particles with momenta $p_i, \, 1\leq i \leq 4$. To describe the external kinematics, we will either use dual-space coordinates (i.e. $p_i=x_{i+1}-x_i$ with $1\leq i \leq 4$)
or momentum-twistor notation \cite{Hodges:2009hk}. We  define 
\begin{equation}
    \label{normalized scattering amplitude}
    {\cal M} := \frac{{\cal A}}{{\cal A}_{\rm tree}} \,,
\end{equation}
where ${\cal A}$ is the color-ordered maximally helicity-violating (MHV) scattering amplitude and ${\cal A}_{\rm tree}$ is its corresponding tree-level value. We will refer to ${\cal M}$ as the scattering amplitude, for simplicity. 
We define the $L$-loop integrand ${\cal I}_L$ for ${\cal M}$
as\footnote{The Amplituhedron is defined for integer dimensions, i.e. $D=4$ and $D=3$ in the ${\cal N}=4$ SYM and ABJM cases, respectively. When considering the amplitude, an infrared regulator needs to be specified, for example dimensional regularization. Since we consider finite quantities, considering the integer-dimensional Amplituhedron integrands is sufficient for our purposes.}
\begin{equation}
\label{amplitude integrand}
{\cal M} \bigg|_{\rm L \, loops} := \left( \prod_{j=5}^{4+L} \int \frac{d^D x_j}{i\pi^{D/2}} \right) \, {\cal I}_L \,,
\end{equation}
where $x_5,x_6, \dots, x_{4+L}$ describe the loop variables. Similarly, we take the $L$-loop integrand ${\cal L}_L$ for the logarithm of the scattering amplitude to be defined as
\begin{equation}
\label{log of amplitude conventions}
\log {\cal M} \bigg|_{\rm L \, loops} := \left( \prod_{j=5}^{4+L} \int \frac{d^D x_j}{i\pi^{D/2}} \right) \, {\cal L}_L \,.
\end{equation}

We now turn to the computation of the integrands ${\cal I}_L$ and ${\cal L}_L$. In order to introduce the main ideas, let us focus first on the ${\cal N}=4$ sYM theory. 
There are many ways of obtaining four-point integrands at high loop orders, including generalized unitarity \cite{Bern:2012uc}, on-shell recursion relations \cite{Arkani-Hamed:2010zjl}, soft-collinear consistency conditions \cite{Bourjaily:2011hi,Bourjaily:2016evz}, and a connection to correlation functions \cite{Eden:2012tu},
for example.
A conceptual breakthrough was achieved in \cite{Arkani-Hamed:2013jha}, where it was proposed that in the ${\cal N}=4$ sYM theory the integrands ${\cal I}_L$ are proportional to the canonical form of a positive geometry known as the \textit{Amplituhedron}. To be more precise, let us take $Z^I_a, \, a=1, \dots, 4$ to be the four-dimensional momentum-twistors that describe the external kinematic data of the scattering process, and let us consider the region in momentum-twistor space described by the constraint
\begin{equation}
    \label{amplituhedron constraint 1}
    \langle 1234 \rangle >0 \,,
\end{equation}
with $\langle 1234 \rangle=\epsilon_{IJKL} Z^I_1 Z^J_2 Z^K_3 Z^L_4$. Moreover, we shall take $L$ lines $l_5:=AB$, $l_6:=CD$, $l_7:=EF$, $\dots$, in momentum-twistor space such that for each one of them we impose
\begin{align}
    \label{amplituhedron constraint 2}
    \langle l_i 12 \rangle &>0 ,\, \quad \langle l_i 23 \rangle >0 ,\, \quad \langle l_i 34 \rangle >0 ,\, \quad \langle l_i 14 \rangle >0 \,, \\
    \label{amplituhedron constraint 3}
    \langle l_i 13 \rangle &<0 ,\, \quad \langle l_i 24 \rangle <0 \,.
\end{align}
Finally, let us demand that each pair of different lines satisfies the mutual positivity constraint 
\begin{equation}
    \label{amplituhedron constraint 4}
    \langle l_i l_j \rangle >0 \,.
\end{equation}
Then, the four-particle $L$-loop MHV Amplituhedron is defined as the set of points in momentum-twistor space that are subjected to the constraints given in \eqref{amplituhedron constraint 1}-\eqref{amplituhedron constraint 4}. One can associate to the Amplituhedron a unique canonical differential form $\Omega$ with logarithmic singularities on the boundaries of the space. Let us introduce the notation
\begin{equation}
    \Omega = \sum_{L=1}^{\infty} \lambda^L \, \Omega_L \,,
\end{equation}
where $\lambda$ is the 't Hooft coupling (we are working on the planar limit). 
Then, the $L$-loop integrand ${\cal I}_L$ for the scattering amplitude is simply given as 
\begin{equation}
\label{normalization I_L}
{\cal I}_L= n_L \, \Omega_L \,,
\end{equation}
where $n_L$ is a normalization factor which we discuss in the Appendix \ref{app: normalization}.

In the following it will prove useful to take into account the pictorial representation introduced in \cite{Arkani-Hamed:2021iya} to describe positive geometries. We will use a node to indicate a one-loop amplituhedron associated to a certain loop variable, i.e. a geometry satisfying the constraints \eqref{amplituhedron constraint 1}-\eqref{amplituhedron constraint 3}, and a dashed light-blue line to describe a mutual positivity condition between a pair of loop variables. As an example, the four-loop amplituhedron will be drawn as
\begin{equation}
\includegraphics[scale=0.10]{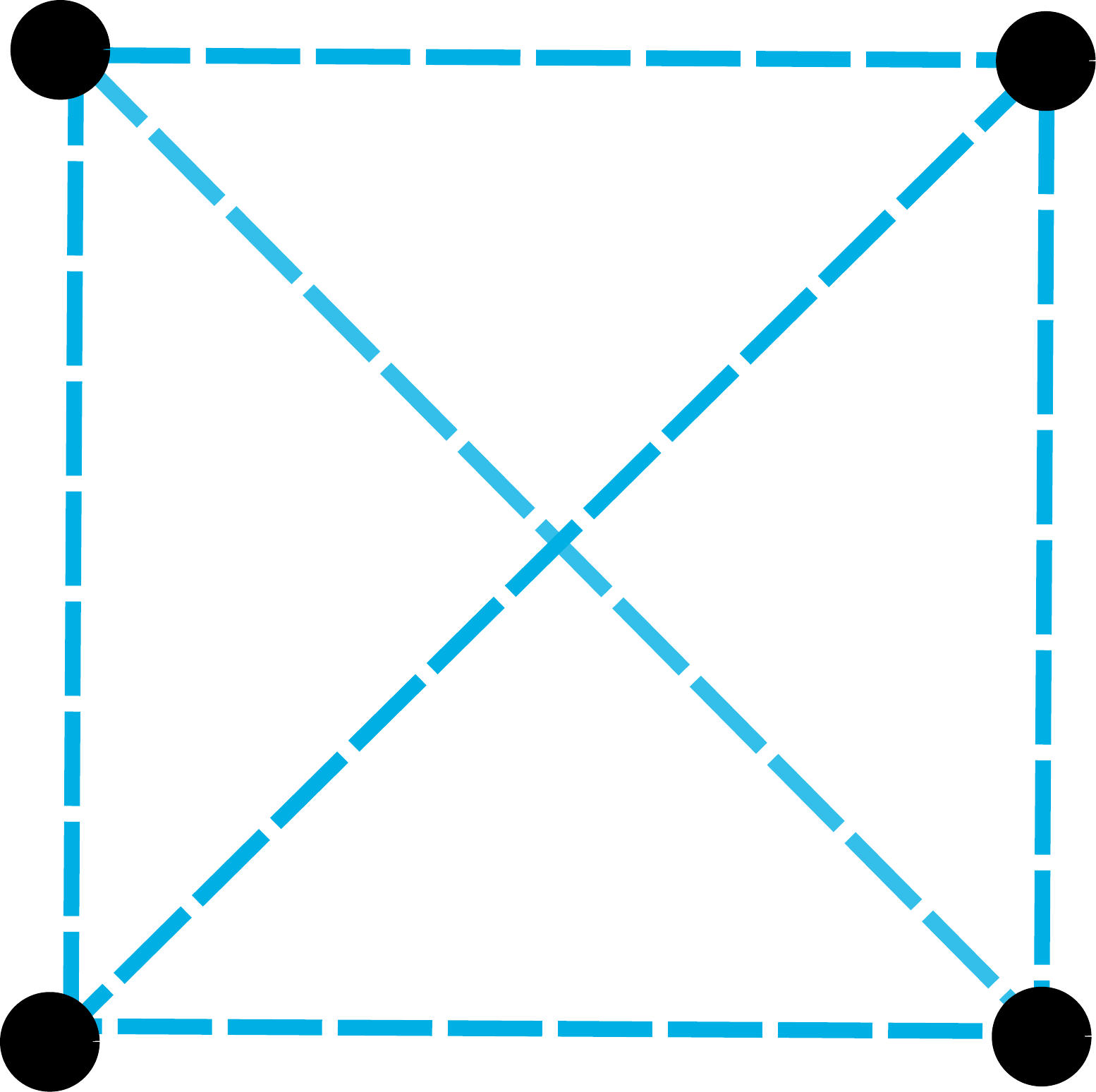} 
\label{four-loop positive geometry}
\end{equation}
As pointed out in \cite{Arkani-Hamed:2021iya}, it turns out to be very convenient to consider also mutual negativity conditions between loop variables. That is, constraints given by
\begin{equation}
    \label{mutual negativity}
    \langle l_i l_j \rangle <0 \,,
\end{equation}
for which we will use thick red lines, e.g. 
\begin{equation}
\includegraphics[scale=0.10]{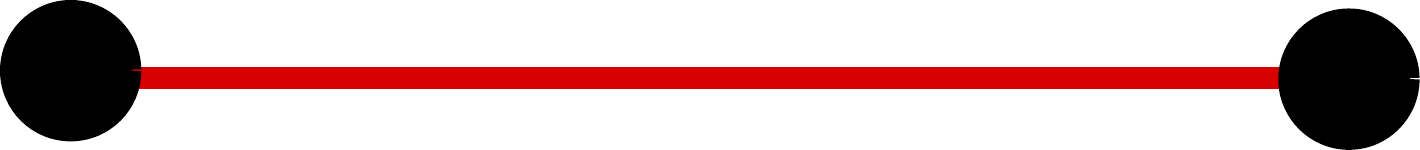} \label{example_neg_geom}
\end{equation}
In order to understand the advantages of using negative geometries for the computation of integrands, let us introduce the notation 
\begin{equation}
    \label{log of positive geometries}
    \tilde{\Omega}:= \log \Omega \,,
\end{equation}
with 
\begin{equation}
    \tilde{\Omega} = \sum_{L=1}^{\infty} \lambda^L \, \tilde{\Omega}_L \,.
\end{equation}
Then, as described in \cite{Arkani-Hamed:2021iya}, one can expand $\tilde{\Omega}$ in terms of \textit{connected} negative geometries. More precisely, 
\begin{equation}
\includegraphics[scale=0.23]{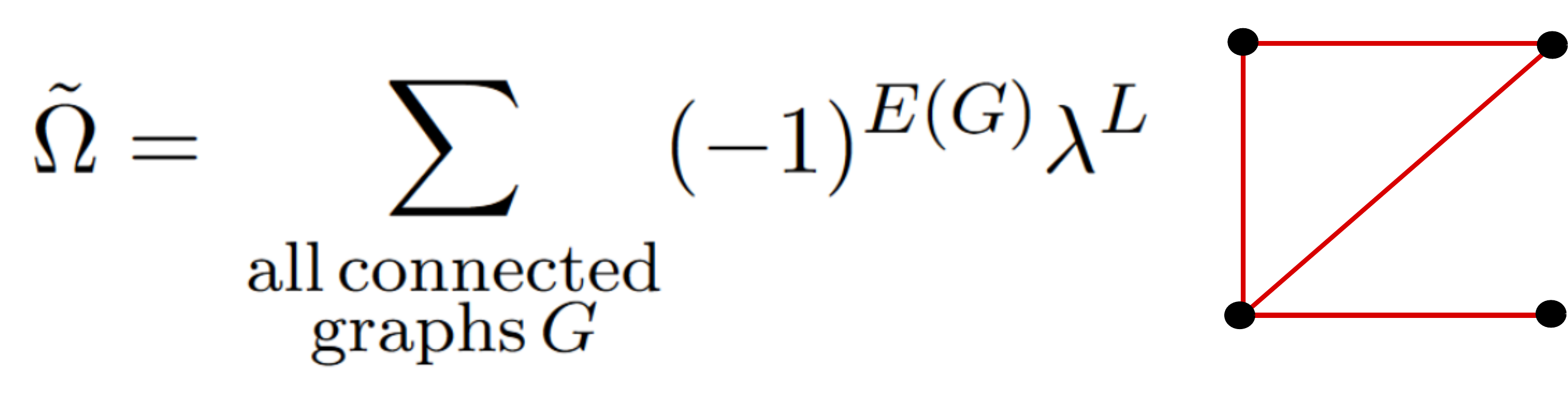} \label{expansion connected neg geom}
\end{equation}
where $E(G)$ is the number of edges of a graph $G$ and $L$ is the corresponding number of vertices. Therefore,
\begin{equation} 
\includegraphics[scale=0.3]{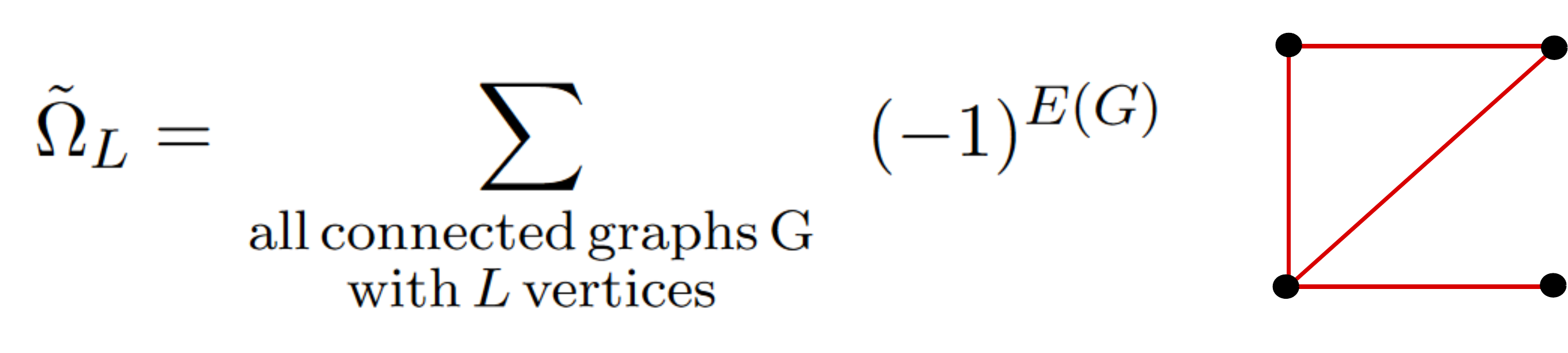} 
\label{expansion tilde Omega-connected graphs}
\end{equation}
The integrand ${\cal L}_L$ is obtained from $\tilde{\Omega}_L$ as
\begin{equation}
\label{normalization L_L}
{\cal L}_L= \tilde{n}_L \, \tilde{\Omega}_L \,.
\end{equation}
We refer again to the Appendix \ref{app: normalization} for the discussion on the computation of the relative normalizations $\tilde{n}_L$.

\subsection{Projected amplituhedron for the ABJM theory}
\label{sec:ABJM-hedron}

Following the ideas of \cite{He:2022cup}, we will now discuss how the previous concepts generalize to the three-dimensional ABJM theory. The four-particle amplituhedron of the ABJM theory can be obtained from projecting the amplituhedron of the ${\cal N}=4$ sYM theory to three dimensions by means of a symplectic constraint. More specifically, the corresponding positive geometry is defined by considering, in addition to the conditions given in \eqref{amplituhedron constraint 1}-\eqref{amplituhedron constraint 4}, the constraints 
\begin{equation}
\label{constraint tree level}
\Sigma_{IJ} Z_i^I Z_{i+1}^J=0 \,,
\end{equation}
for the external kinematic data and
\begin{equation}
\label{L loop constraints}
\Sigma_{IJ} A^I B^J=0 \,,
\end{equation} 
for the loop variables, with $\Sigma$ being a symplectic matrix given as
\begin{equation}
\label{Sigma}
\Sigma= \left( 
\begin{array}{cc} 
0 & \epsilon_{2 \times 2} \\
\epsilon_{2 \times 2} & 0
\end{array}
 \right) \,,
\end{equation}
where $\epsilon_{2 \times 2}$ is a totally anti-symmetric tensor.

One major simplification occurs in ABJM when considering the expansion of $\tilde{\Omega}_L$ into negative geometries, namely that only \textit{bipartite} (connected) graphs are required \cite{He:2022cup}. The latter are defined as those graphs where, after assigning an orientation to each edge, each node is either a sink or a source. Examples of bipartite graphs are
\begin{center}
\includegraphics[scale=0.3]{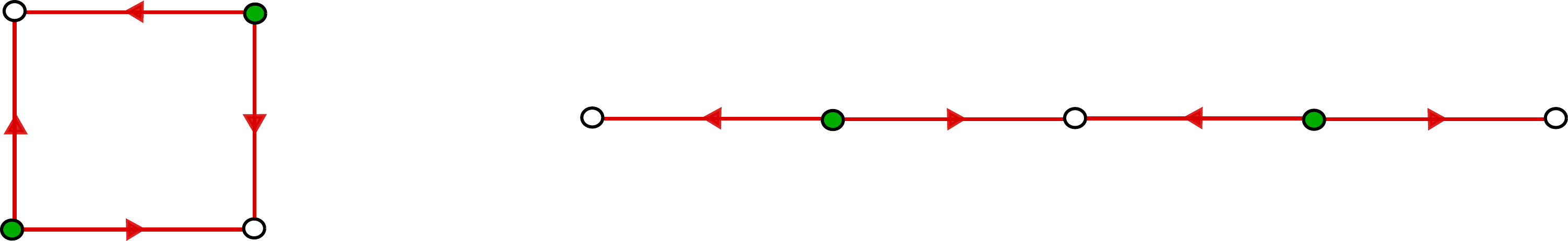}
\end{center}
where green nodes represent sources and white nodes correspond to sinks.
Taking into account the above simplification, the expansion \eqref{expansion tilde Omega-connected graphs} now becomes
\begin{equation}
\includegraphics[scale=0.3]{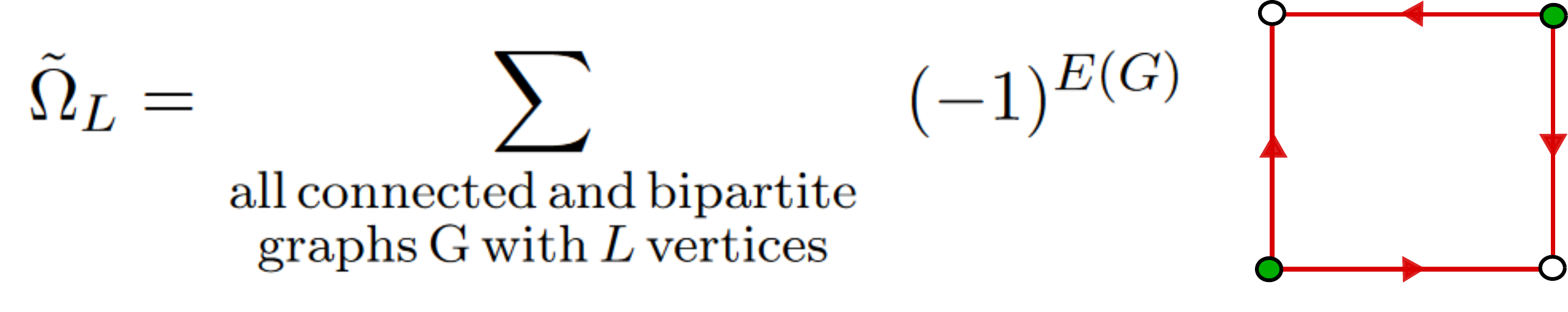} \label{bipartite expansion}
\end{equation}
Using \eqref{bipartite expansion}, the canonical forms $\tilde{\Omega}_L$ were computed in \cite{He:2022cup} up to $L=5$. In particular, for the first three loop orders one gets
\begin{equation}
    \label{amplituhedron canonical forms}
\includegraphics[scale=0.27]{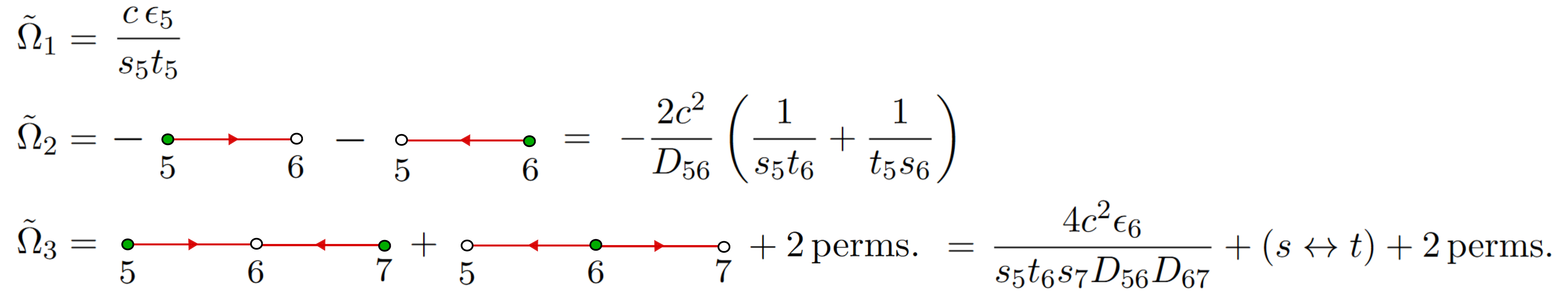}
\end{equation}
with
\begin{align}
\label{s,t,c,D,epsilon definitions}
s_i &:= \langle l_i 12 \rangle \langle l_i 34 \rangle \,,  \quad \quad
t_i := \langle l_i 23 \rangle \langle l_i 14 \rangle \,, \qquad \qquad \qquad
D_{ij} := -\langle l_i l_j \rangle \,, \\
c &:= \langle 1234 \rangle \,, \qquad \qquad
\epsilon_i := \sqrt{\langle l_i 13 \rangle \langle l_i 24 \rangle \langle 1234 \rangle} \,,
\end{align}
and where we are again using the notation $l_5:=AB$, $l_6:=CD$, $l_7:=EF$, $\dots$, for the loop lines, with the permutations being over all nonequivalent configurations of these variables. Let us note that for simplicity of notation we are omitting the $d^3l_i$ factors in the differential forms.

Finally, as shown in Appendix \ref{app: normalization}, the relative normalizations $\tilde{n}_L$ defined in \eqref{normalization L_L} are given by
\begin{equation}
    \label{relative normalizations}
    \tilde{n}_1 = \frac{i}{2\sqrt{\pi}} \,, \qquad \qquad \tilde{n}_2 = \frac{\tilde{n}_1^2}{2!} \,,  \qquad \qquad \tilde{n}_3 = \frac{\tilde{n}_1^3}{3!} \,.
\end{equation}
Let us note that in order to get \eqref{relative normalizations} we are assuming the standard convention
\begin{equation}
    \label{t Hooft coupling}
    \lambda := \frac{N}{k}
\end{equation}
for the 't Hooft coupling of ABJM, with $N$ being the number of colors and $k$ the Chern-Simons level. In the following we are going to use the differential forms \eqref{amplituhedron canonical forms} along with the normalizations \eqref{relative normalizations} as the starting point for performing the loop integrations.

\subsection{Constraints from dual conformal invariance}

Before discussing the explicit integration of the negative geometries, let us analyze the constraints that dual conformal invariance imposes on the integrated expressions. This symmetry can be understood as a consequence of the duality between scattering amplitudes and Wilson loops. The latter is conjectured to hold in ${\cal N}=4$ sYM and to partially extend to the ABJM case \cite{Alday:2007hr,Drummond:2007aua,Brandhuber:2007yx,Drummond:2007cf,Henn:2010ps,Bianchi:2011dg,Chen:2011vv,Leoni:2015zxa}. Indeed, the dual conformal invariance of scattering amplitudes is simply the conformal invariance of the Wilson loops in the dual picture. We will return to the Wilson loops/scattering amplitudes duality in Section \ref{sec: cusp anomalous dimension}.

We will begin again by reviewing the ${\cal N}=4$ sYM case. In four dimensions, the dual conformal invariance of the logarithm of the amplitude implies that there exists a function ${\cal F}_{L-1}$ such that\cite{Alday:2011ga}
\begin{equation}
\label{n=4 general expression}
\left( \prod_{j=6}^{4+L} \int \frac{d^4 x_j}{i\pi^{2}} \right) \, {\cal L}_L = \frac{x_{13}^2 x_{24}^2}{x_{15}^2 x_{25}^2 x_{35}^2 x_{45}^2} \, \frac{{\cal F}_{L-1} \left( z \right)}{\pi^2} \,,
\end{equation}
where the cross-ratio $z$, defined as
\begin{equation}
z= \frac{x_{25}^2 x_{45}^2 x_{13}^2}{x_{15}^2 x_{35}^2 x_{24}^2} \,,
\end{equation}
is the only dual conformally invariant cross-ratio that can be built using the external kinematic data and the unintegrated loop variable $x_5$.
The function ${\cal F}_{L-1}(z)$ has been computed in the literature up to $L=4$ \cite{Alday:2012hy,Alday:2013ip,Henn:2019swt}. Moreover, the above analysis has also been extended to higher-point scattering processes, in which the integrated results depend on functions of more than one cross-ratio. In particular, the five-particle case was studied up to $L=3$ \cite{Chicherin:2022zxo}, while for an arbitrary number of particles the current threshold is $L=2$ \cite{Chicherin:2022bov}.

Let us consider now the above ideas within the context of the ABJM theory. Interestingly, in the three-dimensional case the expression \eqref{n=4 general expression} is incomplete. To see this, it is convenient to use five-dimensional notation to describe the coordinates of the dual space (see for example \cite{Chen:2011vv,Caron-Huot:2012sos} and Appendix \ref{app: five-dimensional notation}).
One can then see that the most general expression that one can construct in order to generalize \eqref{n=4 general expression} to the three-dimensional case is
\begin{equation}
\label{n=4 general expression D=3}
\left( \prod_{j=6}^{4+L} \int \frac{d^3 X_j}{i\pi^{3/2}} \right) \, {\cal L}_L = \left( \frac{X_{13}^2 X_{24}^2}{X_{15}^2 X_{25}^2 X_{35}^2 X_{45}^2} \right)^{\frac{3}{4}} \frac{{\cal F}_{L-1}\left( z \right)}{\sqrt{\pi}}  + \frac{i \epsilon\left( 1,2,3,4,5 \right)}{X_{15}^2 X_{25}^2 X_{35}^2 X_{45}^2} \, \frac{{\cal G}_{L-1} \left( z \right)}{\sqrt{\pi}} \,,
\end{equation}
where capital letters refer to five-dimensional coordinates and 
\begin{equation}
\epsilon(1,2,3,4,5):=\epsilon_{\mu \nu \rho \sigma \eta} X_1^{\mu} X_2^{\nu} X_3^{\rho} X_4^{\sigma} X_5^{\eta} \,.
\end{equation}
Therefore, when going to three dimensions we have to include in \eqref{n=4 general expression} an additional parity-odd term given by a function ${\cal G}_{L-1}(z)$. This is analogous to what was found in ${\cal N}=4$ sYM for the five-particle case \cite{Chicherin:2022zxo}. We will discuss the computation of ${\cal F}(z)$ and ${\cal G}(z)$ in the next section.

%%%%%%%%%%%%%%%%%%%%%%%%%%%%%%%%%%%%%%%%%%%%%%%%%%%%%%%%%%%%%%%%%%%%%%%%%%%%%%%%%%%%%%%%%%%%%%%%%%%%%%%%%%%%%%%%%%%%%%%%%%%%%%%%%%%%%%%%%%%%%%%%%%%%%%%%%%%%%%%%%%%%%%%%%%%%%%
\section{Perturbative analysis}
\label{sec: perturvative results}

We will turn now to the explicit  $L-1$ loop integrations (up to $L=3$) of the negative geometries obtained from the projected amplituhedron for ABJM \cite{He:2022cup}, seeking for the perturbative expansion of the ${\cal F}$ and ${\cal G}$ functions defined in \eqref{n=4 general expression D=3}.

\subsection{Tree level}

Let us begin by computing the tree-level values of ${\cal F}(z)$ and ${\cal G}(z)$. From \eqref{amplituhedron canonical forms} we have

\begin{equation}
\label{Omega tilde 1}
\tilde{\Omega}_1
= 
\frac{\langle 1234 \rangle^{3/2} \sqrt{\langle l_5 13 \rangle \langle l_5 24 \rangle }}{\langle l_512 \rangle \langle l_5 23 \rangle \langle l_5 34 \rangle \langle l_5 14 \rangle} .
\end{equation}
Using the Schouten identity 
\begin{equation}
\label{13 24 twistor identity}
\langle l_5 13 \rangle \langle l_5 24 \rangle= \langle l_5 12 \rangle \langle l_5 34\rangle + \langle l_5 23 \rangle \langle l_5 14\rangle \,.
\end{equation}
we can rewrite \eqref{Omega tilde 1} in terms of five-dimensional dual coordinates (for a nice discussion on that see for example \cite{Arkani-Hamed:2010pyv}) as follows,
\begin{align}
\label{Omega1new}
\tilde{\Omega}_1 = \frac{\sqrt{ X_{13}^2 X_{24}^2 \, (X_{24}^2 X_{15}^2 X_{35}^2+X_{13}^2 X_{25}^2 X_{45}^2)}}{X_{15}^2 X_{25}^2 X_{35}^2 X_{45}^2}
\,.
\end{align}
In order to compute ${\cal F}_0(z)$ and ${\cal G}_0(z)$ it is important to note that 
the integration is over the three-dimensional Minkowski space.
However, \eqref{Omega1new} was derived 
within
the Amplituhedron region defined in \eqref{amplituhedron constraint 1}-\eqref{amplituhedron constraint 4},
and therefore we need to extend its definition.
Indeed, one can see that 
naively
integrating \eqref{Omega1new} over the whole kinematic space 
gives a non-zero result,
in contradiction with 
what is expected for the one-loop four-particle amplitude of ABJM\cite{Agarwal:2008pu,Bianchi:2011fc}. 
This 
issue can be resolved by taking into account the identity\footnote{Any possible overall sign ambiguity that could arise when using \eqref{epsilon identity} should be absorbed in the sign of the overall normalization of the amplitude.}
\begin{equation}
\label{epsilon identity}
 \epsilon(1,2,3,4,5)= -\frac{1}{4} \sqrt{X_{13}^2 X_{24}^2 \, (X_{24}^2 X_{15}^2 X_{35}^2+X_{13}^2 X_{25}^2 X_{45}^2)} \,,
\end{equation}
to analytically continue to Minkowski space. Then, taking into account the normalization presented in \eqref{relative normalizations} and using \eqref{epsilon identity} to rewrite \eqref{Omega1new} in terms of the five-dimensional Levi-Civita tensor we arrive at
\begin{equation}
\label{L1}
{\cal L}_1  = - \frac{2i}{\sqrt{\pi}} \, \frac{\epsilon(1,2,3,4,5)}{X_{15}^2 X_{25}^2 X_{35}^2 X_{45}^2} \, .
\end{equation}
Therefore, comparing to the definition in \eqref{n=4 general expression D=3}, we conclude
\begin{equation}
\label{F0, G0}
{\cal F}_0(z) = 0\,, \quad \text{and} \quad {\cal G}_0(z) = -2 \,.
\end{equation}

\subsection{One loop}
As discussed in the previous sections, the integrand ${\cal L}_2$ that is obtained from the canonical form $\tilde{\Omega}_2$ given in \eqref{amplituhedron canonical forms} reads 
\begin{equation}
\label{L2}
\begin{aligned}
{\cal L}_2
&= \frac{c^2}{4 \pi \, D_{56}} \left( \frac{1}{s_5 t_6} + \frac{1}{t_5 s_6} \right) = \\
 & = -\frac{X_{13}^2 X_{24}^2}{4 \pi \, X_{56}^2} \left( \frac{1}{X_{15}^2 X_{26}^2 X_{35}^2  X_{46}^2} + \frac{1}{X_{16}^2 X_{25}^2 X_{36}^2  X_{45}^2} \right) \,.
\end{aligned} 
\end{equation}
To obtain ${\cal F}_1(z)$ and ${\cal G}_1(z)$ we should now perform one of the loop integrations, which we choose to be the one over $X_6$ (i.e. we take $X_5$ to be the frozen loop variable). This integral turns out to be a triangle integral with three massive legs. Consequently, using the results of Appendix \ref{app:useful_integrals}, we arrive at
\begin{equation}
\label{two loop form amplituhedron-v2-2}
\begin{aligned}
\int \frac{d^3 X_6}{i \pi^{3/2}} \, {\cal L}_2 &= -\frac{X_{13}^2 X_{24}^2}{4 \pi} \, \int \frac{d^3 X_6}{i \pi^{3/2}} \, \frac{1}{ X_{56}^2} \left( \frac{1}{X_{15}^2 X_{26}^2 X_{35}^2  X_{46}^2} + \frac{1}{X_{16}^2 X_{25}^2 X_{36}^2  X_{45}^2} \right) \\
&= - \frac{\sqrt{\pi}}{4} \left( \frac{X_{13}^2 X_{24}^2}{X_{15}^2 X_{25}^2 X_{35}^2 X_{45}^2} \right)^{3/4} \left( z^{1/4} + \frac{1}{z^{1/4}} \right)\,.
\end{aligned}
\end{equation}
Therefore, from \eqref{n=4 general expression D=3} we deduce
\begin{equation}
\label{F1, G1}
{\cal F}_1 (z)= -\frac{\pi}{4} \left( z^{1/4} + \frac{1}{z^{1/4}} \right) \,, \quad\ \text{and} \;\;\, {\cal G}_1 (z)=0 \,.
\end{equation}

\subsection{Two loops}
The integrand ${\cal L}_3$, which can be obtained from \eqref{amplituhedron canonical forms} and \eqref{relative normalizations}, is
\begin{equation}
\label{L3}
{\cal L}_3 
= - \frac{i}{12 \pi^{3/2}} \, \frac{c^2 \epsilon_6}{s_5 t_6 s_7 D_{56} D_{67}} + (s \leftrightarrow t) + 2 \, \text{perms}. 
\end{equation}
In order to compute ${\cal F}_2$ and ${\cal G}_2$ we will perform two of the loop integrations. We will choose to integrate over $X_6$ and $X_7$, and again we will keep $X_5$ frozen. Let us begin by considering the first term of the r.h.s in \eqref{L3}, 
which is explicitly given by
\begin{equation}
\label{first term two loops}
\begin{aligned}
{\cal L}_3^{(1)} 
= \frac{i}{3\pi^{3/2}} \, \frac{X_{13}^2 \, \epsilon(1,2,3,4,6)}{X_{15}^2 X_{35}^2 X_{26}^2 X_{46}^2 X_{17}^2 X_{37}^2 X_{56}^2 X_{67}^2} \,.
\end{aligned}
\end{equation}
First, the integral over $X_7$ is again a triangle integral; therefore, we get
\begin{equation}
\label{first term two loops-2}
\int \frac{d^3X_7}{i \pi^{3/2}} \, {\cal L}_3^{(1)} = \frac{i X_{13} \, \epsilon(1,2,3,4,6)}{3 X_{15}^2 X_{35}^2 X_{26}^2 X_{46}^2 X_{56}^2 X_{16} X_{36}} \,.
\end{equation}
We will turn now to the integration over $X_6$. To compute this integral we will make use of the results derived in Appendix \ref{app:useful_integrals}. In particular, we have 
\begin{equation}
\label{epsilon num integral}
\int \frac{d^3 X_6}{i\pi^{3/2}} \, \frac{\epsilon(1,2,3,4,6)}{X_{26}^2 X_{46}^2  X_{56}^2 X_{16} X_{36}}  = \frac{\epsilon(1,2,3,4,5)}{\left( X_{15}^2 X_{25}^2 X_{35}^2 X_{45}^2 X_{24}^2 \right)^{1/2}} \, \frac{{\cal H}(z)}{\sqrt{\pi z}} \,,
\end{equation}
where the weight-two function ${\cal H}(z)$ takes the form
\begin{equation}
\label{H}
\begin{aligned}
{\cal H}(z) &= \sqrt{\frac{z}{1+z}} \bigg( \pi^2  + 2 \, \log \left( \sqrt{z}+\sqrt{1+z} \right) \log(4z) \\
& \qquad  \qquad \quad \, \, +\text{Li}_2 \left[-2 \left(z+ \sqrt{z(1+z)} \right)  \right] - \text{Li}_2 \left[ -2 \left(z- \sqrt{z(1+z)} \right) \right] \bigg) \,.
\end{aligned}
\end{equation}
Consequently,
\begin{equation}
\label{first term two loops-3}
\int \frac{d^3X_6}{i\pi^{3/2}} \int \frac{d^3X_7}{i\pi^{3/2}} \, {\cal L}_3^{(1)} = \frac{i}{3\sqrt{\pi}} \, \frac{\epsilon(1,2,3,4,5)}{X_{15}^{2} X_{25}^2 X_{35}^{2} X_{45}^2} \, {\cal H}(z) \,.
\end{equation}
For the $X_6 \leftrightarrow X_7$ permutation, i.e. for 
\begin{equation}
\label{second term two loops}
\begin{aligned}
{\cal L}_3^{(2)} := \frac{i}{3\pi^{3/2}} \, \frac{X_{13}^2 \, \epsilon(1,2,3,4,7)}{X_{15}^2 X_{35}^2 X_{27}^2 X_{47}^2 X_{16}^2 X_{36}^2 X_{57}^2 X_{67}^2} \, ,
\end{aligned}
\end{equation}
we similarly get
\begin{equation}
\label{second term two loops-2}
\int \frac{d^3X_6}{i\pi^{3/2}} \int \frac{d^3X_7}{i\pi^{3/2}} \, {\cal L}_3^{(2)} = \frac{i}{3\sqrt{\pi}} \, \frac{\epsilon(1,2,3,4,5)}{X_{15}^{2} X_{25}^2 X_{35}^{2} X_{45}^2} \, {\cal H}(z) \,.
\end{equation}
Finally, let us consider the $X_5 \leftrightarrow X_6$ permutation. That is, let us take
\begin{equation}
\label{third term two loops}
\begin{aligned}
{\cal L}_3^{(3)} := \frac{i}{3\pi^{3/2}} \, \frac{X_{13}^2 \, \epsilon(1,2,3,4,5)}{X_{16}^2 X_{36}^2 X_{25}^2 X_{45}^2 X_{17}^2 X_{37}^2 X_{56}^2 X_{67}^2} \,.
\end{aligned}
\end{equation}
We can see that the integrals of ${\cal L}_3^{(3)}$ over $X_6$ and $X_7$ are simply two triangle integrals, and therefore
\begin{equation}
\label{third term two loops-2}
\begin{aligned}
\int \frac{d^3X_6}{i\pi^{3/2}} \int \frac{d^3X_7}{i\pi^{3/2}} \, {\cal L}_3^{(3)} &=  \frac{i\pi^{3/2}}{3} \, \frac{ \epsilon(1,2,3,4,5)}{X_{15}^2 X_{26}^2 X_{35}^2 X_{45}^2} \,.\\
\end{aligned}
\end{equation}
Adding the corresponding $(s \leftrightarrow t)$ terms, we finally get
\begin{align}
\label{F2, G2}
{\cal F}_2(z)&=0 \,, \nonumber\\
{\cal G}_2(z)&=\frac{2}{3} \left[ {\cal H}(z) + {\cal H} \left( \frac{1}{z} \right) + \pi^2  \right] \,.
\end{align}

%%%%%%%%%%%%%%%%%%%%%%%%%%%%%%%%%%%%%%%%%%%%%%%%%%%%%%%%%%%%%%%%%%%%%%%%%%%%%%%%%%%%%%%%%%%%%%%%%%%%%%%%%%%%%%%%%%%%%%%%%%%%%%%%%%%%%%%%%%%%%%%%%%%%%%%%%%%
\section{Cusp anomalous dimension}
\label{sec: cusp anomalous dimension}

We will discuss now how to obtain the cusp anomalous dimension $\Gamma_{\rm cusp}$ of ABJM from ${\cal F}(z)$ and ${\cal G}(z)$. Let us being then by recalling that $\Gamma_{\rm cusp}$ is defined as \cite{Korchemskaya:1992je}
\begin{equation}
\label{Gamma cusp definition}
\Gamma_{\rm cusp}= \mu \frac{d \log  Z_{\rm cusp}}{d\mu} \,,
\end{equation} 
where $\mu$ is the renormalization scale of the theory and $Z_{\rm cusp}$ is the renormalization factor introduced to renormalize the vacuum expectation value of Wilson loops with light-like cusps.

In ${\cal N}=4$ sYM, an all-loop prediction for $\Gamma_{\rm cusp}$ was obtained following integrability ideas \cite{Beisert:2006ez}. This result was expressed as a function of an interpolating function $h(\lambda)$, which governs the dispersion relation of magnons in the integrability picture\cite{Beisert:2006qh,Minahan:2008hf,Minahan:2009te,Bak:2008vd,Minahan:2009aq,Minahan:2009wg,Leoni:2010tb,Gromov:2014eha,Cavaglia:2016ide}. At weak coupling, it was shown that
\begin{equation}
\label{f n=4}
\Gamma_{\rm cusp}(h)= 4 h^2 - \frac{4}{3} \pi^2 h^4 + \frac{44}{45} \pi^4 h^6 + \dots \,.
\end{equation}
Crucially, in ${\cal N}=4$ sYM the interpolating function was proven to simply be
\begin{equation}
\label{h n=4}
h(\lambda)=\frac{\sqrt{\lambda}}{4\pi} \,,
\end{equation}
at all loops \cite{Gromov:2012eu}.

In terms of the ${\cal N}=4$ sYM result, the cusp anomalous dimension of ABJM was proposed to be\cite{Gromov:2008qe}
\begin{equation}
\label{Gamma cusp proposal ABJM}
 \Gamma_{\rm cusp}^{\rm ABJM}= \frac{1}{4} \Gamma_{\rm cusp}^{{\cal N}=4}\bigg|_{h^{{\cal N}=4} \to h^{\rm ABJM}} \,.
\end{equation}
However, the interpolating function $h(\lambda)$ of ABJM has proven to be much less trivial than its ${\cal N}=4$ sYM counterpart. An all-loop proposal was made in \cite{Gromov:2014eha,Cavaglia:2016ide}, giving
\begin{equation}
\label{h conjecture ABJM}
\lambda= \frac{\sinh(2 \pi h)}{2\pi} \, _3F_2 \left( \frac{1}{2},\frac{1}{2},\frac{1}{2} ;1, \frac{3}{2}; - \sinh^2 (2\pi h) \right) \,.
\end{equation}
Therefore, in the weak-coupling limit we get
\begin{equation}
\label{Gamma cusp conjecture ABJM-weak coupling-2}
\Gamma_{\rm cusp} (\lambda)= \lambda^2 - \pi^2 \lambda^4 + \frac{49 \pi^4}{30} \lambda^6 + \dots \,.
\end{equation}
The above proposal is consistent with the leading-order perturbative result computed in \cite{Griguolo:2012iq}.

Wilson loops seem to be intimately related to scattering amplitudes within the context of the AdS/CFT correspondence. First observed in ${\cal N}=4$ sYM and then partially extended to ABJM, there is a duality that relates scattering amplitudes and polygonal light-like Wilson loops\cite{Alday:2007hr,Drummond:2007aua,Brandhuber:2007yx,Drummond:2007cf,Henn:2010ps,Bianchi:2011dg,Chen:2011vv,Leoni:2015zxa}. To be more specific, let us focus again on a four-particle MHV scattering process characterized by four points $x_i$ in the dual coordinate space. Moreover, let us consider a tetragonal light-like Wilson loop $W_4$ whose vertices locate at the $x_i$ points. Then, the duality identifies
\begin{equation}
    \label{WL SA duality}
    \log {\cal M} \sim \log \langle W_4 \rangle \,,
\end{equation}
at the level of the integrands\footnote{To be precise one should specify a prescription to compute the integrands. At the amplitude's side of the duality the integrand is fixed by using dual-space coordinates and requiring the correct pole structure, while at the Wilson loop's side one should use the method of the lagrangian insertions to build the integrand \cite{Eden:2010zz}.}. We should note that, while in ${\cal N}=4$ sYM the duality is believed to hold for arbitrary number of particles, in ABJM the duality has only been observed for the four-particle case and has been proven to fail for higher numbers of particles\cite{Henn:2010ps,Bianchi:2011dg,Chen:2011vv,Leoni:2015zxa}. 

In order to exploit the Wilson loops/scattering amplitudes duality to relate $\Gamma_{\rm cusp}$ to negative geometries we should recall that the renormalization theory of light-like Wilson loops \cite{Korchemskaya:1992je} implies
\begin{equation}
\label{log WL in terms of Gamma cusp}
\log \langle W_4 \rangle= - 2 \sum_{L=1}^{\infty} \frac{\lambda^L \, \Gamma_{\rm cusp}^{(L)}}{(L \epsilon)^2} + \mathcal{O}(1/\epsilon) \,,
\end{equation}
where $\Gamma_{\rm cusp}^{(L)}$ is the $L$-loop coefficient of the cusp anomalous dimension. Therefore, we get
\begin{equation}
\label{Gamma cusp from F and G}
\int \frac{d^{D} X_5}{i\pi^{D/2}} \left[ \left( \frac{X_{13}^2 X_{24}^2}{X_{15}^2 X_{25}^2 X_{35}^2 X_{45}^2} \right)^{\frac{3}{4}} \frac{{\cal F} \left( z \right)}{\sqrt{\pi}} + \frac{i \epsilon\left( 1,2,3,4,5 \right)}{X_{15}^2 X_{25}^2 X_{35}^2 X_{45}^2} \, \frac{{\cal G} \left( z \right)}{\sqrt{\pi}} \right] = - 2 \sum_{L=1}^{\infty} \frac{\lambda^L \, \Gamma_{\rm cusp}^{(L)}}{\epsilon^2} + \mathcal{O}(1/\epsilon) \,,
\end{equation}
with $D=3-2\epsilon$ and where, as in the ${\cal N}=4$ sYM case \cite{Alday:2013ip}, we have included for dimensional reasons an $L^2$ factor in the $L$-loop contribution. Then, after defining
\begin{align}
\label{F functional}
I_{\cal F}[{\cal F}_{L-1}] &:= \left[-\frac{1}{2\sqrt{\pi}} \int  \frac{d^D X_5}{i\pi^{D/2}} \, \left( \frac{X_{13}^2 X_{24}^2}{X_{15}^2 X_{25}^2 X_{35}^2 X_{45}^2} \right)^{\frac{3}{4}} {\cal F}_{L-1} \left( z \right) \right]_{1/\epsilon^2 \, {\rm term}} \,, \\
\label{G functional}
I_{\cal G}[{\cal G}_{L-1}] &:= \left[ -\frac{1}{2\sqrt{\pi}} \int \frac{d^D X_5}{i\pi^{D/2}} \,  \frac{i \epsilon(1,2,3,4,5)}{X_{15}^2 X_{25}^2 X_{35}^2 X_{45}^2} \, {\cal G}_{L-1} \left( z \right) \right]_{1/\epsilon^2 \, {\rm term}}  \,,
\end{align}
we have
\begin{equation}
\label{Gamma cusp from F and G-functionals}
\Gamma_{\rm cusp}^{(L)} = I_{\cal F}[{\cal F}_{L-1}] + I_{\cal G}[{\cal G}_{L-1}] \,.
\end{equation}
As shown in Appendix \ref{app:useful_integrals}, using Feynman parametrization we get
\begin{align}
\label{F functional-zp}
I_{\cal F}[z^p] &= -\frac{2}{\sqrt{\pi} \, \Gamma \left( \frac{3}{4} +p \right) \Gamma \left( \frac{3}{4} -p \right)} \,, \\
\label{G functional-zp}
I_{\cal G}[z^p] &= 0 \,.
\end{align}
Therefore, we see that \eqref{Gamma cusp from F and G-functionals} together with \eqref{F functional-zp} and \eqref{G functional-zp} gives us a prescription to compute $\Gamma_{\rm cusp}$ from the knowledge of ${\cal F}(z)$. In particular, the results of Section \ref{sec: perturvative results} allow us to recover the leading-order contribution to the cusp anomalous dimension of ABJM, i.e.
\begin{equation}
    \label{Gamma cusp result}
    \Gamma_{\rm cusp} (\lambda)= \lambda^2 + \mathcal{O}(\lambda^4)\,,
\end{equation}
in accordance with eq. \eqref{Gamma cusp conjecture ABJM-weak coupling-2}.

%%%%%%%%%%%%%%%%%%%%%%%%%%%%%%%%%%%%%%%%%%%%%%%%%%%%%%%%%%%%%%%%%%%%%%%%%%%%%%%%%%%%%%%%%%%%%%%%%%%%%%%%%%%%%%%%%%%%%%%%%%%%%%%%%%%%%%%%%%%%%%%%%%%%%%%%%%%%%%%%%%%%%%%%%%%%%%

\section{Transcendental weight properties of the results}
\label{sec:transcendental}

In this section we discuss the transcendental weight properties of our results, and more generally of loop corrections in the three-dimensional ABJM theory.

\subsection{Preliminaries}

The appearance of series expansions in the coupling constant $\lambda$ or in the dimensional regularization parameter $\epsilon$ is ubiquitous in the context of quantum field theories. In this framework, the analysis of the transcendental degree properties of the different terms in a given expansion has proven to be a powerful tool for the study of scattering amplitudes and Feynman integrals. As an example, the method of canonical differential equations \cite{Henn:2013pwa} relies on insight about which loop integrands integrate to uniform transcendental weight functions. In this section we will turn to the study of the transcendental degree properties of the results we have presented so far.

Let us begin by recalling that the {\it{transcendental weight}} (also called \textit{degree of transcendentality}) $T$ of a function $f$ is defined as the number of iterated integrals that are needed to compute $f$ \cite{Henn:2013pwa}, e.g. $T({\rm Li}_n)=n$. Moreover, one can extend the definition of $T$ to transcendental numbers, i.e. numbers that can not be obtained as the solution of a polynomial equation with rational coefficients. For example, $T(\pi)=1$ and $T(\zeta_n)=n$. A series expansion is said to have uniform degree of transcendentality (often abbreviated as UT) when all its terms have the same degree $T$. 
Moreover, when discussing Laurent expansions in the dimensional regularization parameter $\epsilon$, it is natural to assign weight $-1$ to $\epsilon$ (see \cite{Henn:2020omi} for a review).

\subsection{Transcendental weight properties of three-dimensional scattering amplitudes and Feynman integrals in the literature}

Scattering amplitudes in ${\cal N}=4$ sYM are conjectured to have uniform transcendental weight \cite{Bern:2005iz,Dixon:2011pw,Arkani-Hamed:2012zlh,Kotikov:2006ts,Beisert:2006ez} (and, as a consequence, the same holds for the cusp anomalous dimension $\Gamma_{\rm cusp}$). 
Furthermore, it has been observed that the leading transcendental weight terms of $\Gamma_{\rm cusp}$ agree between ${\cal N}=4$ sYM and QCD \cite{Kotikov:2002ab,Kotikov:2004er}. 
The specific weight at a given loop order $L$ depends on the choice of the effective coupling constant. In ${\cal N}=4$ sYM, a natural normalization choice for the effective coupling is $g^2 := g_{\rm YM}^2/(16 \pi^2)$. With that normalization choice, the $L$-loop coefficients of scattering amplitudes are observed to have weight $2L$. See \cite{Henn:2020omi} for more details.

Similar uniform weight properties of scattering amplitudes and $\Gamma_{\rm cusp}$ have been observed in ABJM \cite{Chen:2011vv,Bianchi:2013pva,Caron-Huot:2012sos,He:2022lfz,Bianchi:2012cq,Gromov:2008qe}. 
Since the specific weight at each loop order depends on the choice of effective coupling (and whether its weight is counted or not), let us recall that the standard choice used in the literature is
\begin{align}\label{eq:couplingstandardABJM}
    \lambda = \frac{N}{k}\,.
\end{align}
With the choice (\ref{eq:couplingstandardABJM}) of the effective coupling, we find that the results presented in the literature for scattering amplitudes and Wilson loops are consistent with $L$-loop coefficients having weight $L$. 
This is also the case for the conjectured all-loop formula for the cusp anomalous dimension (see eq. \eqref{Gamma cusp conjecture ABJM-weak coupling-2}) when multiplied by $1/\epsilon^2$ (recalling that it appears in scattering amplitudes in this combination).

Let us note that we could alternatively use the following effective coupling,
\begin{align}\label{lambdaeff}
    \tilde{\lambda} = \frac{\lambda}{\sqrt{\pi}}\,.
\end{align}
This would lead to a transcendental weight of $3L/2$ at $L$ loops. This may be a natural choice, as in this case one could say that the weight is $D_0 L/2$ at $L$ loops, with $D=D_{0}-2 \epsilon$, which then applies both to ${\cal N}=4$ sYM and ABJM.

Ultimately, the transcendental weight properties of amplitudes can be traced back to properties of Feynman loop integrals. 
Here we wish to remind readers of what is known, and to establish  what our conventions for loop integrals are. 
When computing Feynman integrals, the following  measure is commonly used, \begin{align}\label{eq:loopmeasure}
    \frac{d^{D}k}{i \pi^{D/2}}\,,
\end{align} for each loop integration.
This convention has the effect that when switching to Feynman parametrization there are no explicit factors of $\pi$.
Of course, when computing QFT observables the choice of measure cannot be seen in an isolated way, but is related to the choice of effective coupling, such as eq. \eqref{eq:couplingstandardABJM} or eq. (\ref{lambdaeff}).

It has been observed that with the convention (\ref{eq:loopmeasure}) the maximal weight of $L$-loop Feynman integrals is $D_{0} L/2$ if $D_{0}$ is even (again we are taking $D=D_{0}-2 \epsilon$). 
This is in agreement with \cite{Hannesdottir:2021kpd}.
For example, the well-known four-point amplitude in ${\cal N}=4$ SYM is given by $g^2$ times the following box integral (for $D=4-2\epsilon$), 
\begin{align}
    \int \frac{d^{D}k}{i \pi^{D/2}} \frac{s t }{(k-p_1)^2 k^2 (k+p_2)^2 (k+p_2+p_3)^2} \,,
\end{align}
which has uniform weight $2$, in agreement the statements made above.

Much less is known about integrals in odd space-time dimensions $D_{0}$. It is interesting to inspect the integrals computed in this paper, see Section \ref{sec: perturvative results} and Appendix \ref{app:useful_integrals}. Using the integration measure (\ref{eq:loopmeasure}) and the alternative convention \eqref{lambdaeff} for the coupling constant\footnote{With the alternative convention $\tilde{\lambda} = \frac{\lambda}{\sqrt{\pi}}$ one gets that the normalizations $\tilde{n}_L$ become 
$$
n_1=i  \,, \qquad n_2=\tilde{n}_2=\frac{\tilde{n}_1^2}{2!} \,, \qquad n_3=\tilde{n}_3=\frac{\tilde{n}_1^3}{3!} \,.
$$
This change in the normalization has to be taken into account when revisiting the integrals \eqref{two loop form amplituhedron-v2-2} and \eqref{first term two loops-3}.
} we find that 
the one-loop integral \eqref{two loop form amplituhedron-v2-2} (see also the triangle integral in \eqref{triangle integral-1} and the epsilon integral in \eqref{integral 2 loops-epsilon numerator-1}) and the two-loop integral \eqref{first term two loops-3} have weight $D_0 L/2$. This appears to lie within the bound proposed by \cite{Hannesdottir:2021kpd}.

\subsection{Transcendental weight properties of ${\cal F}$ and ${\cal{G}}$.}

Let us now turn to our tree-level, one-loop  and two-loop  results for the functions ${\cal F}$ and ${\cal G}$, given in eqs. (\ref{F0, G0}), (\ref{F1, G1}) and (\ref{F2, G2}), respectively.
Putting them together, we have 
\begin{align}
    {\cal F}(z) &= \sum_{L=1}^{\infty} {\cal F}_{L-1}(z) \lambda^L \,, \\
    {\cal G}(z) &= \sum_{L=1}^{\infty} {\cal G}_{L-1}(z) \lambda^L \,,
\end{align}
with
\begin{align}
    \label{F results}
    \frac{{\cal F}(z)}{\lambda} &= -\frac{\pi}{4} \left( z^{1/4} + \frac{1}{z^{1/4}} \right) \lambda + \mathcal{O} \left( \lambda^3 \right) \,, \\
    \label{G results}
    \frac{{\cal G}(z)}{\lambda} &= -2 + \frac{2}{3} \left[ {\cal H}(z) + {\cal H} \left( \frac{1}{z} \right) + \pi^2  \right] \lambda^2 + \mathcal{O} \left( \lambda^3 \right) \,.
\end{align}
We see that the coefficients 
of the $\lambda^L$ powers have weight $L$,
in agreement with the discussion of the previous subsection.
Equivalently, when using the effective coupling $\tilde{\lambda}$ from eq. (\ref{lambdaeff}) one would find weight $3L/2$ at $L$ loops.

Finally, let us comment on the function space we found.
This is best analyzed by the
 \textit{symbol} \cite{Goncharov:2010jf},
 which is an important concept related to a transcendental function. Let us consider a transcendental function $f$ of weight $n$ whose total derivative can be writen as
\begin{equation}
    \label{symbol def-1}
    df= \sum_i g_i \, d\log \omega_i \,,
\end{equation}
where the $g_i$ are functions of weight $n-1$ and the $\omega_i$ are rational functions called \textit{letters}. The set of all letters of a transcendental function is known as its \textit{alphabet}. Then, the symbol of $f$ is defined recursively as
\begin{equation}
    \label{symbol def-2}
    {\cal S}(f)= \sum_i {\cal S}(g_i) \otimes \omega_i \,.
\end{equation}
The knowledge of the symbol of a transcendental function, combined with other information, is often enough to bootstrap the result of the corresponding iterated integral. As an example, this program has been used to compute several scattering amplitudes in ${\cal N}=4$ sYM\cite{Dixon:2011pw,Dixon:2011nj,Dixon:2014voa,Dixon:2015iva,Caron-Huot:2016owq,Drummond:2014ffa,Dixon:2016nkn}. With this into consideration, obtaining the alphabet of the results presented in Section \ref{sec: perturvative results} could open the door for a bootstrap computation of higher-loop terms.

It is therefore interesting to determine the alphabet of letters of our two-loop functions, given in eq. \eqref{H}.
It can be readily read off using the definitions (\ref{symbol def-1}) and (\ref{symbol def-2}).
In terms of the variables $z$ one finds that the letters have a square root dependence. The latter can be removed by changing variables as follows,
\begin{align}
z = \frac{4 q^2}{(1-q^2)^2} \,,
\end{align}
with $0<q<1$.
Using this variable, we find the symbol 
\begin{align}
    {\cal S}\left( \sqrt{\frac{z}{1+z}} H(z) \right) = \frac{q^2}{(1-q^2)^2} \otimes \frac{1+q}{1-q}  \,.
    \end{align}
In other words, the letters that compose the alphabet at two loops are
\begin{align}
\vec{\omega} = \{q , 1-q, 1+q \} \,.
\end{align}

%%%%%%%%%%%%%%%%%%%%%%%%%%%%%%%%%%%%%%%%%%%%%%%%%%%%%%%%%%%%%%%%%%%%%%%%%%%%%%%%%%%%%%%%%%%%%%%%%%%%%%%%%%%%%%%%%%%%%%%%%%%%%%%%%%%%%%%%%%%%%%%%%%%%%%%%%%%%%%%%%%%%%%%%%%%%%%

\section{Conformal invariance of leading singularities}
\label{sec: Conformal invariance of leading singularities}

In this section we study the symmetry properties of the leading singularities that 
characterize the integrated negative geometries of ABJM, as previously explored in \cite{Chicherin:2022bov,Chicherin:2022zxo} for the ${\cal N}=4$ sYM case. We begin our analysis by a short review of the four-dimensional results, and then we turn to the discussion of the conformal invariance of the three-dimensional leading singularities. To that end, we discuss separately the parity-even and -odd terms that appear in \eqref{n=4 general expression D=3} after the integration of the negative geometries.

\subsection{Review of four-dimensional results}

Let us start by reviewing the hidden conformal symmetry that was observed in ${\cal N}=4$ sYM at tree level. In this case one has 
\begin{equation}
\label{d4 tree level}
{\cal L}_1= - \frac{x_{13}^2 x_{24}^2}{x_{15}^2 x_{25}^2 x_{35}^2 x_{45}^2}\,.
\end{equation}
More generally, in the generic $L$-loop case one can write
\begin{equation}
\label{eq:leading_singularities_def}
\left( \prod_{j=6}^{4+L} \int \frac{d^4 x_j}{i\pi^2} \right) \, {\cal L}_L = \sum_{i=1}^k R_{L-1,i}\, T_{L-1,i}\,,
\end{equation}
where $k$ is some integer, $T_{L-1,i}$ are transcendental functions, and $R_{L-1,i}$ are rational functions known as \textit{leading singularities}. As an example, when applying the definition \eqref{eq:leading_singularities_def} to \eqref{d4 tree level} we get
\begin{equation}
\label{d4 tree level LS and T}
R_{0}= \frac{x_{13}^2 x_{24}^2}{x_{15}^2 x_{25}^2 x_{35}^2 x_{45}^2} \,, \quad \text{and} \quad T_{0}=-1 \,.
\end{equation}

At this point it is useful to take advantage of the conformal covariance of the l.h.s. of \eqref{eq:leading_singularities_def}, which allows us to go to the frame at which $x_5 \to \infty$. The convenience of this frame relies on the fact that now we can write all functions using four-particle kinematic notation. To be more precise, let us define the leading singularities $r_{L,i}$ in the $x_5 \to \infty$ frame as
\begin{equation}
\label{eq:x5_inf_limit}
r_{L,i} := \lim_{x_5 \to \infty} \, (x_5^2)^4 \, R_{L,i}\,.
\end{equation} 
Then, at tree level we get
\begin{equation}
\label{eq:r0}
r_0= x_{13}^2 x_{24}^2= s \,t \,,
\end{equation}
where $s:=(p_1+p_2)^2=x_{13}^2$ and $t:=(p_1+p_4)^2=x_{24}^2$ are the well-known Mandelstam variables. Moreover, in terms of four-dimensional spinor-helicity variables we have
\begin{equation}
\label{r0-spinor helicity}
r_0= \langle 12 \rangle \langle 14 \rangle \left[ 12 \right] \left[ 14 \right]\,,
\end{equation}
where $\langle ij \rangle= \epsilon_{ab} \,\lambda^a_i \lambda^b_j $ and $\left[ ij \right]= \epsilon_{\dot{a} \dot{b}} \,\tilde{\lambda}^{\dot{a}}_i \tilde{\lambda}^{\dot{b}}_j$, and with the spinor-helicity variables $\lambda^a$ and $\tilde{\lambda}^{\dot{a}}$ being defined as
\begin{equation}
\label{4d spinor helicity}
p^{a\dot{a}} = p_{\mu} \left(\sigma^{\mu} \right)^{a\dot{a}} = \lambda^a \tilde{\lambda}^{\dot{a}} \,.
\end{equation}

Now, in order to discuss the conformal invariance of the leading singularities defined above let us recall that in four-particle kinematics the generator of special conformal transformations is written as
\begin{equation}
\label{4d K}
K_{a\dot{a}}=\sum_{i=1}^4 \frac{\partial^2}{\partial \lambda^a \partial \tilde{\lambda}^{\dot{a}}}\,.
\end{equation}
Therefore, we can see that the leading singularity \eqref{r0-spinor helicity} is not invariant under special conformal transformations. Instead, as observed in \cite{Chicherin:2022bov}, in order to get a conformally invariant quantity one should multiply the leading singularity by the Parke-Taylor factor ${\rm PT}$, defined as
\begin{equation}
\label{PT 4d}
{\rm PT}=\frac{1}{\langle 12 \rangle \langle 23 \rangle \langle 34 \rangle \langle 41 \rangle} \,.
\end{equation}
That is, when normalizing the leading singularity $r_0$ as
\begin{equation}
\hat{r}_0 := {\rm PT}\, r_0= \frac{\left[ 12 \right] \left[ 41 \right]}{\langle 23 \rangle \langle 34 \rangle}\,,
\end{equation}
one gets a conformally invariant function. Finally, we should note that, as shown in \cite{Chicherin:2022bov}, these results generalize to higher-point tree-level leading singularities.

\subsection{Leading singularities of parity-even terms}

In order to generalize the above results to the ABJM case, let us first review how the conformal generators look when written in terms of three-dimensional kinematic variables. First, let us introduce three-dimensional spinor-helicity variables as
\begin{equation}
\label{3d spinor helicity}
p^{ab}=\lambda^a \lambda^b \,,
\end{equation}
with
\begin{equation}
\label{3d momentum matrix}
p^{ab}=\left( \sigma^{\mu} \right)^{ab} p_{\mu}= \left( \begin{array}{cc}
p^0-p^1 & p^2 \\
p^2 & p^0+p^1
\end{array}
\right)\,.
\end{equation}
Moreover, let us define the Mandelstam variables $s_{ij}$ as
\begin{equation}
\label{sij spinor helicity}
s_{ij} := (p_i + p_j)^2= -\langle ij \rangle^2\,.
\end{equation}
Then, one can write the conformal generators of the one-particle representation of the $\mathfrak{osp}(6|4)$ superalgebra of ABJM \cite{Bargheer:2010hn} as 
\begin{align*}
\label{conformal generators 3d}
P^{ab} &= \lambda^a \lambda^b\,, \qquad \qquad
L^a_b = \lambda^a \partial_b-\frac{1}{2} \delta^a_b \lambda^c \partial_c \,, \\
K_{ab} &= \partial_a \partial_b \,, \qquad \qquad D = \frac{1}{2} \lambda^a \partial_a + \frac{1}{2}\,,
\end{align*}
were $L^a_b$ are the generators of rotations, $D$ is the dilatation operator, and $K_{ab}$ is the generator of special conformal transformations. As for multi-particle representations, one can construct the generators by adding up the corresponding single-particle operators. In particular, for the four-particle case the three-dimensional generalization of \eqref{4d K} reads
\begin{equation}
\label{eq:3d_special_conformal_generator}
K_{ab}=\sum_{i=1}^4 \partial_a^i \partial_b^i \,.
\end{equation}

We can turn now to the symmetry analysis of the three-dimensional leading singularities. We should recall that in \eqref{n=4 general expression D=3} we found that in ABJM the result of performing $L-1$ loop integrations over the $L$-loop integrand ${\cal L}_L$ can be separated into parity-even and -odd terms. In order to discuss the conformal properties of the integrated geometries, we will find instructive to study those terms separately. 

Let us begin by considering the parity-even terms ${\cal P}_e$. In section \ref{sec: perturvative results} we found that these terms were given by
\begin{equation}
\label{parity even terms}
\frac{{\cal P}_e}{\lambda} = -\frac{\sqrt{\pi}}{4} \left( \frac{X_{13}^2 X_{24}^2}{X_{15}^2 X_{25}^2 X_{35}^2 X_{45}^2} \right)^{3/4} \left( z^{1/4} + \frac{1}{z^{1/4}} \right) \lambda + \mathcal{O}(\lambda^3) \,.
\end{equation} 
Therefore, at one-loop order the leading singularities are
\begin{equation*}
\label{even leading sing}
\begin{aligned}
\qquad R_{e,1} = \left( \frac{X_{13}^2 X_{24}^2}{X_{15}^2 X_{25}^2 X_{35}^2 X_{45}^2} \right)^{3/4} \,  z^{1/4}\, \,, \qquad  \text{and} \qquad 
R_{e,2} = \left( \frac{X_{13}^2 X_{24}^2}{X_{15}^2 X_{25}^2 X_{35}^2 X_{45}^2} \right)^{3/4} \,  \frac{1}{z^{1/4}}\,. \\
\end{aligned}
\end{equation*}
As in the ${\cal N}=4$ sYM case, we will take the $x_5 \to \infty$ limit and we will define
\begin{equation}
\label{x5 inf LS}
r_{e,i} := \lim_{x_5 \to \infty} (x_5^2)^3 \, R_{e,i} \,,
\end{equation}
such that
\begin{equation*}
\label{even leading sing-2}
\begin{aligned}
\qquad \, \, \, r_{e,1} &= s \sqrt{t} \,,   \qquad \text{and} \qquad 
r_{e,2} &= t \sqrt{s} \,. \\
\end{aligned}
\end{equation*}
When going to spinor-helicity notation we have to be careful with the sign that comes from taking the square root in \eqref{sij spinor helicity}. 
However, given that a constant overall sign in the leading singularities is not important when discussing their symmetry properties, from now on we will ignore it, simply assuming a plus sign\footnote{We will also ignore overall factors of $i$.}. We remind the reader that the sign could be different depending on the kinematic region, and therefore our conclusions regarding conformal invariance will only be valid locally. Then, we can write
\begin{equation}
\label{even leading sing-3}
\begin{aligned}
\qquad \, \, \, r_{e,1} &= \langle 12 \rangle^2 \langle 14 \rangle \,, \qquad \text{and} \qquad 
r_{e,2} = \langle 12 \rangle \langle 14 \rangle^2 \,.
\end{aligned}
\end{equation}

As a first test, let us consider what happens when we multiply \eqref{even leading sing-3} by the Parke-Taylor factor given in \eqref{PT 4d}. We get
\begin{equation}
\label{even leading sing-4}
\begin{aligned}
{\rm PT} \, r_{e,1} &=  \frac{\langle 12 \rangle}{\langle 23 \rangle \langle 34 \rangle \langle 14 \rangle} \,, \qquad \quad \text{and} \qquad
{\rm PT} \, r_{e,2} = \frac{\langle 14 \rangle}{\langle 12 \rangle \langle 23 \rangle \langle 34 \rangle}\,,  \\
\end{aligned}
\end{equation}
which are not invariant under the action of the special conformal generators given in \eqref{eq:3d_special_conformal_generator}. In order to understand why \eqref{even leading sing-4} fails to be conformally invariant we should recall in ${\cal N}=4$ sYM the Parke-Taylor factor appears within the tree-level amplitudes as
\begin{equation}
\label{tree level n=4}
{\cal A}^{{\cal N}=4 \, {\rm sYM}}_4 = {\rm PT} \, \delta^{(4)}(P) \, \delta^{(8)}(Q)\,.
\end{equation}
Comparing with tree-level amplitudes in ABJM, which are given by \cite{Bargheer:2010hn}
\begin{equation}
\label{tree level abjm}
{\cal A}^{{\rm ABJM}}_4 = -\frac{\delta^{(3)}(P) \, \delta^{(6)}(Q)}{\langle 12 \rangle \langle 14 \rangle}\,,
\end{equation}
we find natural to define a three-dimensional Parke-Taylor factor as
\begin{equation}
\label{parke taylor 3d}
{\rm PT}^{(3)} := \frac{1}{\langle 12 \rangle \langle 14 \rangle}\,.
\end{equation}
and to normalize the three-dimensional leading singularities as
\begin{equation}
\label{renormalized LS}
\hat{r}= {\rm PT}^{(3)} \, r \,.
\end{equation} 
Indeed, with this normalization we get
\begin{equation}
\label{even leading sing-5}
\begin{aligned}
\hat{r}_{e,1} &= \sqrt{s} =  \langle 12 \rangle \,, \qquad \text{and} \qquad 
\hat{r}_{e,2} = \sqrt{t} = \langle 14 \rangle\,, \\
\end{aligned}
\end{equation}
which are invariant under the generators \eqref{eq:3d_special_conformal_generator} of special conformal transformations. Therefore, we conclude that the one-loop leading singularities of the parity-even terms of \eqref{n=4 general expression D=3} become conformally invariant when normalized by the three-dimensional Parke-Taylor factor \eqref{parke taylor 3d} and when evaluated in the $x_5 \to \infty$ frame.

\subsection{Leading singularities of parity-odd terms}

Let us turn now to the study of the parity-odd terms ${\cal P}_o$ of \eqref{n=4 general expression D=3}. In order to analyze the conformal invariance of their leading singularities, we must identify first how to take the $x_5 \to \infty$ limit in expressions that include contractions $\epsilon(1,2,3,4,5)$ with the
five-dimensional Levi-Civita tensor. To that end, it is instructive to recall the identity 
\begin{equation}
\label{5D epsilon^2-section 5}
\epsilon(1,2,3,4,5)^2= \frac{x_{13}^2 x_{24}^2}{16} (x_{24}^2 x_{15}^2 x_{35}^2+x_{13}^2 x_{25}^2 x_{45}^2) \,,
\end{equation}
which can be found in the discussion of Appendix \ref{app: five-dimensional notation}. 
As in the previous section, we will ignore the sign ambiguity that comes from taking square roots.
Our equalities should be understood up to a possible overall $\pm i$ factor, and our conclusions about symmetry invariance will only be valid locally. Then, we get\footnote{It is interesting to note that \eqref{5d epsilon limit} can also be obtain from the contraction
\begin{equation}
\label{5d epsilon infty vector}
\epsilon(1,2,3,4,I) = \frac{1}{2} \sqrt{s t (s+t)} \,,
\end{equation}
where $I=(1,0,\vec{0})$ corresponds to a point in infinity\cite{Caron-Huot:2012sos}. That is,
we can alternatively write
\begin{equation}
\label{5d epsilon limit-2}
\lim_{x_5 \to \infty}  \left( x_5^2 \right)^{-1} \, \epsilon(1,2,3,4,5) = \frac{1}{2} \, \epsilon(1,2,3,4,I) \,.
\end{equation}}
\begin{equation}
\label{5d epsilon limit}
\lim_{x_5 \to \infty} \left( x_5^2 \right)^{-1} \, \epsilon(1,2,3,4,5) = \frac{1}{4} \sqrt{s t (s+t)} \,.
\end{equation}

Having discussed how to correctly take the $x_5 \to \infty$ limit, we can now safely turn to the analysis of the symmetry properties of the parity-odd terms ${\cal P}_o$. From \eqref{G results} we have
\begin{equation}
\label{parity odd terms}
\frac{{\cal P}_o}{\lambda} = \frac{i\epsilon(1,2,3,4,5)}{\sqrt{\pi} \, X_{15}^2 X_{25}^2 X_{35}^2 X_{45}^2} \left[ -2 + \frac{2}{3} \left( {\cal H}(z) + {\cal H} \left( \frac{1}{z} \right) + \pi^2 \right) \lambda^2 \right] + \mathcal{O}(\lambda^3) \,.
\end{equation} 
Then, up to the loop order we have studied we get that the leading singularities are
\begin{equation*}
\label{parity odd ls}
\begin{aligned}
R_{o,1} &= \frac{4 \, \epsilon(1,2,3,4,5)}{X_{15}^2 X_{25}^2 X_{35}^2 X_{45}^2} \,,  \\
R_{o,2} &= R_{o,1} \, \sqrt{\frac{z}{1+z}} \,, \\
R_{o,3} &= R_{o,1} \, \sqrt{\frac{1}{1+z}} \,.
\end{aligned}
\end{equation*}
Therefore, in the $x_5 \to \infty$ we get
\begin{equation}
\label{parity odd ls-2}
\begin{aligned}
r_{o,1} &=  \sqrt{s t (s+t)} \,, \\
r_{o,2} &= s \sqrt{t} \,, \\
r_{o,3} &= t \sqrt{s} \,,
\end{aligned}
\end{equation}
and, after normalizing with the three-dimensional Parke-Taylor factor,  we have
\begin{equation}
\label{parity odd ls-3}
\begin{aligned}
\hat{r}_{o,1} &= \sqrt{s+t}= \langle 13 \rangle \,, \\
\hat{r}_{o,2} &= \sqrt{s}= \langle 12 \rangle \,, \\
\hat{r}_{o,3} &= \sqrt{t}= \langle 14 \rangle \,,
\end{aligned}
\end{equation}
We see that the expressions in \eqref{parity odd ls-3} are conformally invariant, in a similar way to what was observed for the parity-even terms.

%%%%%%%%%%%%%%%%%%%%%%%%%%%%%%%%%%%%%%%%%%%%%%%%%%%%%%%%%%%%%%%%%%%%%%%%%%%%%%%%%%%%%%%%%%%%%%%%%%%%%%%%%%%%%%%%%%%%%%%%%%%%%%%%%%%%%%%%%%%%%%%%%%%%%%%%%%%%%%%%%%%%%%%%%%%%%%%%%%%%%%%%%%%%%%%%%%%%%%%%%%%%%%%%%%%%%%%%%%%%%%%%%%%%%%%%%%%%%%%%%%%%%%%%%%%%%%%%%%%%%%%%%%%%%%%%%%%%%%%%%%%%%%%%%%%%%%%%%%%%%%%%%%%%%%%%%%%%%%%%%%%%%%%%%%%%%%%%%%%%%%%%%%%%%%%%%%%%%%%%%%%%%%%%%%%%%%%%%%%%%%%%%%%%%%%%%%%%%%%%%%%%%%%%%%%%%%%%%%%%%%%%%%%%%%%%%%%%%%%%%%%%%%%%%%%%%%%%%%%%%%%%%%%%%%%%%%%%%%%%%%%%%%%%%%%%%%%%%%%%%%%%%%%%%%%%%%%%%%%%%%%%%%%%%%%%%%%%%%%%%%%%%%%%%%%%%%%%%%%%%%%%%%%%%%%%%%%%%%%%%%%%%%%%%%%%%%%%%%%%%%%%%%%%%%%%%%%%%%%%%%%%%%%%%%%%%%%%%%%%%%%%%%%%%%%%%%%%%%%%%%%%%%%%%%%%%%%%%%%%%%%%%%%%%%%%%%%%%%%%%%%%%%%%%%%%%%%%%%%%%%%%%%%%%%%%%%%%%%%%%%%%%%%%%%%%%%%%%%%%%%%%%%%%%%%%%%%%%%%%%%%%%%%%%%%%%%%%%%%%%%%%%%%%%%%%%%%%%%%%%%%%%%%%%%%%%%%%%%%%%%%%%%%%%%%%%%%%%%%%%%%%%%%%%%%%%%%%%%%%%%%%%%%%%%%%%%%%%%%%%%%%%%%%%%%%%%%%%%%%%%%%%%%%%%%%%%%%%%%%%%%%%%%%%%%%%%%%%%%%%%%%%%%%%%%%%%%%%%%%%%%%%%%%%%%%%%%%%%%%%%%%%%%%%%%%%%%%%%%%%%%%%%%%%%%%%%%%%%%%%%%%%%%%%%%%%%%%%%%%%%%%%%%%%%%%%%%%%%%%%%%%%%%%%%%%%%%%%%%%%%%%%%%%%%%%%%%%%%%%%%%%%%%%%%%%%%%%%%%%%%%%%%%%%%%%%%%%%%%%%%%%%%%%%%%%%%%%%%%%%%%%%%%%%%%%%%%%%%%%%%%%%%%%%%%%%%%%%%%%%%%%%%%%%%%%%%%%%%%%%%%%%%%%%%%%%%%%%%%%%%%%%%%%%%%%%%%%%%%%%%%%%%%%%%%%%%%%%%%%%%%%%%%%%%%%%%%%%%%%%%%%%%%%%%%%%%%%%%%%%%%%%%%%%%%%%%%%%%%%%%%%%%%%%%%%%%%%%%%%%%%%%%%%%%%%%%%%%%%%%%%%%%%%%%%%%%%%%%%%%%%%%%%%

\section{Conclusions}
\label{sec: conclusions}

In this paper we have studied, for the three-dimensional ${\cal N}=6$ Chern-Simons-matter (ABJM) theory, the result of performing $L-1$ loop integrations over the $L$-loop integrand of the logarithm of the four-particle scattering amplitude. We have used the negative geometries that come from the projected amplituhedron for the ABJM theory \cite{He:2022cup} as the starting point for constructing the integrands. We have found that the dual conformal symmetry of the amplitudes allows for the presence of both parity-even and -odd terms in the integrated results, in a similar way to what was described for the five-particle case in the ${\cal N}=4$ super Yang-Mills (sYM) theory \cite{Chicherin:2022zxo}. We have performed the explicit integrations up to $L=3$, and we have found that the results are given by infrared-finite quantities
with uniform degree of transcendentality, 
as it was also observed for the analogous quantities in ${\cal N}=4$ sYM.
Moreover, we have constructed functionals that allow one to compute the ABJM cusp anomalous dimension $\Gamma_{\rm cusp}$ using the integrated negative geometries as the input, and by doing so we have recovered the known first non-trivial order of $\Gamma_{\rm cusp}$\cite{Griguolo:2012iq,Gromov:2008qe}. Finally, we have discussed the symmetry properties of the leading singularities associated to the integrated results. We have found that the leading singularities have a hidden conformal symmetry (in the frame in which the unintegrated loop variable goes to infinity, and after normalization with a three-dimensional generalization of the Parke-Taylor factor), 
extending the four-dimensional analysis of \cite{Chicherin:2022bov,Chicherin:2022zxo}. 

There are a number of exciting open questions. In the ${\cal N}=4$ super Yang-Mills theory, a useful dual perspective is provided by the duality between scattering amplitudes and Wilson loops. This allows one to think of loop integrands as derivatives of Wilson loop correlators w.r.t. the coupling. More precisely, the derivatives produce Lagrangian insertions, and it is natural to consider 
\begin{equation}
\frac{\langle W_4 \, L(x_5) \rangle}{\langle W_4 \rangle} \,,
\end{equation}
where $W_4$ is the dual polygonal Wilson loop, $L$ is the Lagrangian of the theory, and $x_5$ is the unintegrated loop variable. 
It would be desirable to extend this to the ABJM case. However, an immediate difficulty is that the Lagrangian in ABJM (and in Chern-Simons theories in general) is not gauge invariant, as the variation of the action includes a non-trivial topological term.

Another interesting direction that arises from our results is the question about their generalization to scattering processes with higher numbers of particles. To that end, one could expect to apply the idea of the projected amplituhedron proposed in \cite{He:2022cup} to compute the corresponding negative geometries for higher multiplicities, to then study the properties of the integrated results as we have performed at four points. Furthermore, in view of the Grassmanian formulas proposed for the ABJM theory \cite{Lee:2010du,Huang:2013owa,Huang:2014xza}, it would be interesting to analyze the symmetry properties of the leading singularities in terms of a Grassmanian formulation, as it was done in \cite{Chicherin:2022bov} for the ${\cal N}=4$ sYM case.

Considering the relation between integrated negative geometries in ${\cal N}=4$ sYM and all-plus amplitudes in the pure Yang Mills theory\cite{Chicherin:2022bov,Chicherin:2022zxo}, one intriguing question to address is a possible generalization of this result to the ABJM case. In this regard, one should take into account that an analogous relation between ABJM and the pure Chern-Simons theory does not seem possible, as the latter is a topological theory and therefore has a vanishing S-matrix. However, it would be interesting to investigate a possible relation between ABJM and less supersymmetric Chern-Simons-matter theories.

Furthermore, it would be interesting to carry on the integrations to higher loop orders. For $L \geq 4$ it seems to be far less trivial how to perform the integrations by first principles. However, many useful methods have been developed over the last years to overcome the difficulties that arise when computing Feynman integrals. In particular, the method of differential equations \cite{Henn:2013pwa} appears as a promising tool to solve the $L=4$ case, which in turn would allow to reproduce the next-to-leading non-trivial order of the ABJM cusp anomalous dimension.

Finally, an interesting problem to investigate is whether one can sum infinite series of negative-geometry diagrams. This question was addressed in $D=4$ in \cite{Arkani-Hamed:2021iya}, where the all-loop sum of \textit{ladder} and \textit{tree} diagrams was performed. The crucial observation used in \cite{Arkani-Hamed:2021iya} is that in $D=4$ the Laplace operator $\Box=\partial^{\mu} \partial_{\mu}$ acts on the propagator as
\begin{equation}
    \label{4d Box on propagator}
    \Box \left( \frac{1}{x^2} \right) = -4i \pi^2 \, \delta^{(4)} (x)\,.
\end{equation}
This allows one to recursively relate diagrams that differ only on one loop integration, and ends up giving second-order differential equations for ${\cal F}_{\rm ladder}(z)$ and ${\cal F}_{\rm tree}(z)$. It is useful to note that eq. \eqref{4d Box on propagator} naturally arises in Fourier space, as in this context the four-dimensional propagator is simply $1/k^2$. Unfortunately,  
in $D=3$ there is a mismatch of dimensions in momentum space, which prevents us from using  the Laplace equation trick.
Nevertheless, we find it likely that the finite integrals have other special properties that may lead to simplifications.
It would be interesting to work further in this direction. An all-loop sum of negative-geometry diagrams could be a first step towards obtaining non-perturbative results for the cusp anomalous dimension $\Gamma_{\rm cusp}$. Moreover, it could set the stage for a non-perturbative computation of the interpolating function $h(\lambda)$, for which all-loop proposals exist \cite{Gromov:2014eha,Cavaglia:2016ide}.

%%%%%%%%%%%%%%%%%%%%%%%%%%%%%%%%%%%%%%%%%%%%%%%%%%%%%%%%%%%%%%%%%%%%%%%%%%%%%%%%%%%%%%%%%%%%%%%%%%%%%%%%%%%%%%%%%%%%%%%%%%%%%%%%%%%%%%%%%%%%%%%%%%%%%%%%%%%%%%%%%%%%%%%%%%%%%%
\section*{Acknowledgements}

We would like to thank Paolo Benincasa, Jungwon Lim, Antonela Matija\v{s}i\'{c}, Julian Miczajka and Emery Sokatchev for useful discussions. 
ML would like to especially thank Diego Correa for insightful comments and supervision. 
ML is supported by fellowships from CONICET (Argentina) and DAAD (Germany).
This research was supported by the Munich Institute for Astro-, Particle and BioPhysics (MIAPbP) which is funded by the Deutsche Forschungsgemeinschaft (DFG, German Research Foundation) under Germany's Excellence Strategy - EXC-2094 - 390783311.
This research received funding from the European Research Council (ERC) under the European Union's Horizon 2020 research and innovation programme (grant agreement No 725110), {\it Novel structures in scattering amplitudes}. 
\\[0.2cm]
\textbf{Note added}\\
After this work was completed we learned that Song He, Chia-Kai Kuo, Zhenjie Li and Yao-Qi Zhang independently obtained similar results \cite{He:2023exb}. We would like to thank them for correspondence and for confirming agreement with our two-loop result, see \eqref{H} and \eqref{F2, G2}.

%%%%%%%%%%%%%%%%%%%%%%%%%%%%%%%%%%%%%%%%%%%%%%%%%%%%%%%%%%%%%%%%%%%%%%%%%%%%%%%%%%%%%%%%%%%%%%%%%%%%%%%%%%%%%%%%%%%%%%%%%%%%%%%%%%%%%%%%%%%%%%%%%%%%%%%%%%%%%%%%%%%%%%%%%%%%%%
\appendix

\section{Five-dimensional notation}
\label{app: five-dimensional notation}

When working with three-dimensional dual-coordinates it turns out useful to consider the embedding of the three-dimensional Minkowski space into the five-dimensional projective light-cone. One of the main advantages of this parametrization lies in the fact it allows to write three-dimensional dual-conformal invariants simply as five-dimensional expressions that respect Lorentz and scale invariance.

To be more precise, let us consider a five-dimensional Minkowski space with $(-,-,+,+,+)$ signature and with coordinates $(X^1,X^2,X^3,X^4,X^5)$. Then the light-cone is defined by the constraint
\begin{equation}
\label{5D light-cone-1}
-(X^1)^2-(X^2)^2+(X^3)^2+(X^4)^2+(X^5)^2=0 \,.
\end{equation}
Let us note that the constraint \eqref{5D light-cone-1} is invariant under a rescaling of the coordinates, and therefore defines a projective space with 3 degrees of freedom, as expected. It is useful to switch to light-cone coordinates $(X^+,X^-,X^2,X^4,X^5)$, with $X^+$ and $X^-$ given by
\begin{equation}
\label{light-cone coordinates}
X^+=\frac{X^1+X^3}{\sqrt{2}} \,, \qquad \text{and} \qquad X^-=\frac{X^1-X^3}{\sqrt{2}} \,,
\end{equation}
so that \eqref{5D light-cone-1} becomes
\begin{equation}
\label{5D light-cone-2}
-2 \, X^+ \, X^- -(X^2)^2+(X^4)^2+(X^5)^2=0 \,.
\end{equation}
The embedding of the three-dimensional Minkowski space\footnote{We are using the $(-,+,+)$ signature for the three-dimensional Minkowski space.} with coordinates $(x^0,x^1,x^2)$ into the five-dimensional space can be defined as
\begin{equation}
\label{embedding}
(X^+,X^-,X^2,X^4,X^5)=\left( \frac{x^{\mu} x_{\mu}}{2},1,x^0,x^1,x^2 \right) \,.
\end{equation}
It is straightforward then to check that \eqref{embedding} satisfies \eqref{5D light-cone-2}. Moreover, under this parametrization we have
\begin{equation}
\label{3D and 5D invariants}
(X_i-X_j)^2=-2 X_i.X_j=(x_i-x_j)^2 \,,
\end{equation}
where $x_i$ and $x_j$ are points in the three-dimensional space and $X_i$ and $X_j$ are their corresponding images under the mapping \eqref{embedding}.

In order to simplify notation, we will write the contraction of the dual coordinates with the five-dimensional Levi-Civita tensor as
\begin{equation}
\label{5D epsilon}
\epsilon(1,2,3,4,5):=\epsilon_{\mu \nu \rho \sigma \eta} X_1^{\mu} X_2^{\nu} X_3^{\rho} X_4^{\sigma} X_5^{\eta} \,.
\end{equation}
Let us recall some properties of \eqref{5D epsilon}. In the first place, one can rewrite \eqref{5D epsilon} in terms of three-dimensional dual-coordinates as\cite{Chen:2011vv}
\begin{equation}
\label{5D epsilon-2}
\epsilon(1,2,3,4,5)= \frac{1}{2} \left( x_{51}^2 \epsilon_{\mu\nu\rho} x_{21}^{\mu} x_{31}^{\nu} x_{41}^{\rho} + x_{31}^2 \epsilon_{\mu\nu\rho} x_{51}^{\mu} x_{21}^{\nu} x_{41}^{\rho} \right) \,.
\end{equation}
Also, the product of two contractions is given by
\begin{equation}
\label{5D epsilon product}
\epsilon(1,2,3,4,5) \, \epsilon(1,2,3,4,6)= \frac{X^4_{13} X^4_{24}}{32} \left( \frac{X^2_{15} X^2_{36}+X^2_{16} X^2_{35}}{X^2_{13}} + \frac{X^2_{25} X^2_{46}+X^2_{26} X^2_{45}}{X^2_{24}} - X^2_{56}\right) \,.
\end{equation}
In particular, we have
\begin{equation}
\label{5D epsilon^2}
\epsilon(1,2,3,4,5)^2= \frac{X_{13}^2 X_{24}^2}{16} (X_{24}^2 X_{15}^2 X_{35}^2+X_{13}^2 X_{25}^2 X_{45}^2) \,.
\end{equation}
Finally, following \cite{Caron-Huot:2012sos} we can define a measure on the five-dimensional light-cone as
\begin{equation}
\label{measure light-cone}
d^3 X := \int \frac{d^5 X}{\text{Vol}[\text{GL(1)}]} \, \delta(X^2) \,,
\end{equation}
where the factor $\delta(X^2)$ is included to satisfy the constraint given in \eqref{5D light-cone-1}, while the denominator $\text{Vol}[\text{GL(1)}]$ eliminates the redundancy coming from the projective invariance of the light-cone. Therefore, we get
\begin{equation}
\label{relation measures}
\int d^3 X \equiv \int d^3 x \,.
\end{equation}

%%%%%%%%%%%%%%%%%%%%%%%%%%%%%%%%%%%%%%%%%%%%%%%%%%%%%%%%%%%%%%%%%%%%%%%%%%%%%%%%%%%%%%%%%%%%%%%%%%%%%%%%%%%%%%%%%%%%%%%%%%%%%%%%%%%%%%%%%%%%%%%%%%%%%%%%%%%%%%%%%%%%%%%%%%%%%%
\section{Normalization of negative geometries}
\label{app: normalization}

As shown in \eqref{normalization I_L} and \eqref{normalization L_L}, there are relative normalizations $n_L$ and $\tilde{n}_L$ between the integrands ${\cal I}_L$ and ${\cal L}_L$ and the canonical forms $\Omega_L$ and $\tilde{\Omega}_L$. We will discuss their computation in this section.

As a first step, we should note that the definitions \eqref{amplitude integrand} and \eqref{log of amplitude conventions} imply
\begin{align}
\label{IL in terms of LL-1}
{\cal I}_1 &= {\cal L}_1 \,, \\
\label{IL in terms of LL-2}
{\cal I}_2 &= {\cal L}_2 + \frac{1}{2} {\cal L}_1^2 \,, \\
\label{IL in terms of LL-3}
{\cal I}_3 &= {\cal L}_3 + {\cal L}_2 \, {\cal L}_1+ \frac{1}{6} {\cal L}_1^3 \,.
\end{align}
On the other hand, from \eqref{expansion tilde Omega-connected graphs} we get
\begin{align}
\label{Omega_L in terms of Tilde_Omega_L-1}
\Omega_1 &= \tilde{\Omega}_1 \,, \\
\label{Omega_L in terms of Tilde_Omega_L-2}
\Omega_2 &= \tilde{\Omega}_2 + \tilde{\Omega}_1^2 \,, \\
\label{Omega_L in terms of Tilde_Omega_L-3}
\Omega_3 &= \tilde{\Omega}_3 +3 \, \tilde{\Omega}_2 \tilde{\Omega}_1 + \tilde{\Omega}_1^3 \,.
\end{align}
Therefore, using the definitions \eqref{normalization I_L} and \eqref{normalization L_L} and the expansions \eqref{IL in terms of LL-1}-\eqref{Omega_L in terms of Tilde_Omega_L-3} we have
\begin{align}
\label{L in terms of omega tilde-1}
{\cal L}_1 &= n_1 \, \tilde{\Omega}_1 \,, \\
\label{L in terms of omega tilde-2}
{\cal L}_2 &=  n_2 \tilde{\Omega}_2 + \left( n_2 - \frac{n_1^2}{2} \right) \tilde{\Omega}_1^2 \,, \\
\label{L in terms of omega tilde-3}
{\cal L}_3 &= n_3 \tilde{\Omega}_3 + \tilde{\Omega}_1 \, \tilde{\Omega}_2 \left( 3n_3 -n_2 n_1 \right)+ \tilde{\Omega}_1^3 \left( n_3 -n_2 n_1 + \frac{n_1^3}{3} \right) \,.
\end{align}

To fix the values of the $n_L$ coefficients we will follow the ideas of \cite{Bourjaily:2011hi}. These authors used the fact that the integrands ${\cal L}_L$ in planar ${\cal N}=4$ sYM should behave as ${\cal O}(1/\delta)$ in the limit 
\begin{equation}
\label{normalization limit}
\langle l_5 12 \rangle \sim \delta \,, \qquad \quad \text{and} \qquad \langle l_523 \rangle \sim \delta \,,
\end{equation}
while all other brackets remain non-vanishing. This property makes sure that infrared divergences exponentiate (after integration). 
A similar analysis can be done in the ABJM case. Therefore, noticing that \eqref{amplituhedron canonical forms} implies\begin{equation}
    \label{tilde omegas in normalization limit} 
 \tilde{\Omega}_L \sim {\cal O} (1/\delta)  \qquad \text{for} \qquad 1 \leq L \leq 3 \,,
\end{equation}
in the limit \eqref{normalization limit}, and demanding the same behaviour for the l.h.s. of \eqref{L in terms of omega tilde-1}-\eqref{L in terms of omega tilde-3}, we get
\begin{equation}
\label{relation between normalization coefficients-2}
n_1=\tilde{n}_1 \,, \qquad n_2=\tilde{n}_2=\frac{\tilde{n}_1^2}{2!} \,, \qquad n_3=\tilde{n}_3=\frac{\tilde{n}_1^3}{3!} \,.
\end{equation}
Finally, comparing the explicit formulas for ${\cal I}_1$ and ${\cal I}_2$ given in \cite{Bianchi:2011dg,Chen:2011vv} to the expressions for $\Omega_1$ and $\Omega_2$ obtained from the results of \cite{He:2022cup} we get
\begin{equation}
\label{tilde_n_1}
\tilde{n}_1= \frac{i}{2\sqrt{\pi}} \,.
\end{equation}

%%%%%%%%%%%%%%%%%%%%%%%%%%%%%%%%%%%%%%%%%%%%%%%%%%%%%%%%%%%%%%%%%%%%%%%%%%%%%%%%%%%%%%%%%%%%%%%%%%%%%%%%%%%%%%%%%%%%%%%%%%%%%%%%%%%%%%%%%%%%%%%%%%%%%%%%%%%%%%%%%%%%%%%%%%%%%%
\section{Useful integrals}
\label{app:useful_integrals}

We present here several useful integrals for computing the perturbative results of Section \ref{sec: perturvative results}, as well as the integrals that give us the functionals $I_{\cal F}$ and $I_{\cal G}$ in Section \ref{sec: cusp anomalous dimension}.

\subsection{Triangle integral}
Let us begin with a triangle integral in three dimensions and with three massive legs. This integral first appears in the one-loop analysis in \eqref{two loop form amplituhedron-v2-2}, and it is explicitly given by
\begin{equation}
\label{triangle integral-1}
{\cal T} := \int \frac{d^3X_6}{i \pi^{3/2}} \, \frac{1}{X_{26}^2 X_{46}^2 X_{56}^2} \,.
\end{equation}
Using the standard Feynman parametrization one gets 
\begin{equation}
\label{triangle integral-2}
{\cal T} = \frac{\pi^{3/2}}{\sqrt{X_{25}^2 X_{45}^2 X_{24}^2}} \,.
\end{equation}
It is interesting to note that the functional form of the result (\ref{triangle integral-2}) can also be obtained from noticing that the integral \eqref{triangle integral-1} has dual conformal invariance.

\subsection{Five-leg integral with an epsilon numerator}

Let us consider now the integral
\begin{equation}
\label{integral 2 loops-epsilon numerator-1}
{\cal E} := \int \frac{d^3 X_6}{i\pi^{3/2}} \, \frac{\epsilon(1,2,3,4,6)}{X_{26}^2 X_{46}^2  X_{56}^2 X_{16} X_{36}} \,.
\end{equation}
which shows up at the two-loop computation in \eqref{epsilon num integral}. Introducing Feynman parameters, we have
\begin{equation}
\label{integral 2 loops-epsilon numerator-2}
\begin{aligned}
{\cal E} &=\frac{\epsilon_{\mu \nu \rho \sigma \eta} X_1^{\mu} X_2^{\nu} X_3^{\rho} X_4^{\sigma}}{\pi \, \text{Vol}[\text{GL(1)}]}  \left( \prod_{i=1}^5 \int_{0}^{\infty} d\alpha_i \, \right) \, (\alpha_1 \alpha_3)^{-1/2} \, \partial^{\eta}_Y \left[ \int \frac{d^3 X_6}{i\pi^{3/2}} \, \frac{1}{\left( -2Y.X_6 \right)^3} \right] \,,
\end{aligned}
\end{equation}
where we have defined
\begin{equation}
\label{Y 5D}
Y:= \sum_{i=1}^5 \alpha_i \, X_i \,,
\end{equation}
and we have used \eqref{3D and 5D invariants}. Then, performing the space-time integral we have
\begin{equation}
\label{integral 2 loops-epsilon numerator-3}
\begin{aligned}
{\cal E} &=  \frac{3}{4 \sqrt{\pi} \, \text{Vol}[\text{GL(1)}]} \left( \prod_{i=1}^5 \int_{0}^{\infty} d\alpha_i \, \right) \, \frac{(\alpha_1 \alpha_3)^{-1/2} \, \epsilon(1,2,3,4,Y)}{\left(  -Y^2 \right)^{\frac{5}{2}}} \,. \\
\end{aligned}
\end{equation}
At this point is useful to define
\begin{equation}
\label{beta variables}
\beta_i := \alpha_i X_{i5}^2 \quad \text{for} \quad i=1,\dots ,4 \,,
\end{equation}
and to mod out the ${\rm GL}(1)$ invariance by setting
\begin{equation}
\label{constraint betas}
\sum_{i=1}^4 \beta_i=1 \,.
\end{equation}
Then, performing the integral over $\alpha_5$ we get
\begin{equation}
\label{integral 2 loops-epsilon numerator-4}
\begin{aligned}
{\cal E} &= \frac{ \epsilon(1,2,3,4,5)}{\sqrt{\pi} \, \left( X_{15}^2 X_{35}^2 \right)^{1/2} X_{25}^2 X_{45}^2}{} \left( \prod_{i=1}^4 \int_{0}^{\infty} d\beta_i \, \right) \delta \left( \sum_{i=1}^4 \beta_i -1 \right) \, \frac{(\beta_1 \beta_3)^{-1/2}}{\left( \beta_1 \beta_3 \frac{X_{13}^2}{X_{15}^2 X_{35}^2} + \beta_2 \beta_4 \frac{X_{24}^2}{X_{25}^2 X_{45}^2} \right)^{\frac{1}{2}}} \,.
\end{aligned}
\end{equation}
The number of remaining integrals can be further simplified by defining
\begin{equation}
\label{gamma variables}
\begin{aligned}
\beta_1 &:= \gamma_1 \gamma_2 \,, \qquad \qquad \, \, \, \beta_2 := \gamma_1 (1-\gamma_2) \,, \\
\beta_3 &:= (1-\gamma_1) \gamma_3 \,, \qquad \beta_4 := (1-\gamma_1) (1-\gamma_3) \,.
\end{aligned}
\end{equation}
Let us note that the constraint \eqref{constraint betas} is trivially satisfied by the $\gamma$'s. In terms of these variables we get
\begin{equation}
\label{integral 2 loops-epsilon numerator-5}
\begin{aligned}
{\cal E} &= \frac{\epsilon(1,2,3,4,5)}{\left( X_{15}^2 X_{25}^2 X_{35}^2 X_{45}^2 X_{24}^2 \right)^{1/2}} \,  \frac{{\cal H}(z)}{\sqrt{\pi \, z}} \,,
\end{aligned}
\end{equation}
with
\begin{equation}
\label{H-1}
{\cal H} (z) := \sqrt{z} \int_{0}^{1} d\gamma_2 \, \int_{0}^{1} d\gamma_3 \, \frac{(\gamma_2 \gamma_3)^{-1/2}}{\left[ \gamma_2 \gamma_3 z + (1-\gamma_2) (1-\gamma_3) \right]^{\frac{1}{2}}} \,.
\end{equation}
Let us focus on the integral \eqref{H-1}, which as we shall see can be solved by iterating Feynman parametrizations. Making the change of variables
\begin{equation}
\label{first change of variables}
\gamma_2 \to \frac{1}{1+\gamma_2} \, , \qquad \gamma_3 \to \frac{\gamma_3}{1+\gamma_3} \,,
\end{equation}
and introducing Feynman parameters one gets
\begin{equation}
\label{H-2}
{\cal H}(z) = \frac{\sqrt{z}}{\pi} \, \left( \prod_{i=1}^3 \int_0^{\infty} d\eta_i \right) \, \frac{1}{\sqrt{\eta_3}  \, (\eta_1 + \eta_2) (\eta_1+1)  (\eta_2 + \eta_3 + z)} \,.
\end{equation}
Moreover, taking
\begin{equation}
\label{second change of variables}
\eta_3 \to \eta_3^2 \,,
\end{equation}
and making a further Feynman parametrization we have
\begin{equation}
\label{H-3}
{\cal H}(z) =  \int_0^{\infty} d\nu_1 \, \int_0^{\infty} d\nu_2 \,  \frac{\sqrt{z}}{(\nu_1 + \nu_2) (\nu_1 + 1) \sqrt{\nu_2+ z}} \,.
\end{equation}
Finally, defining 
\begin{equation}
\label{theta}
\theta:= \sqrt{\nu_2 +z} \,,
\end{equation}
and integrating over $\theta$ we arrive at
\begin{equation}
\label{H-4}
\begin{aligned}
{\cal H}(z) &= \sqrt{\frac{z}{1+z}} \bigg( \pi^2  + 2 \, \log \left( \sqrt{z}+\sqrt{1+z} \right) \log(4z) \\
& \qquad  \qquad \quad \, \, +\text{Li}_2 \left[-2 \left(z+ \sqrt{z(1+z)} \right)  \right] - \text{Li}_2 \left[ -2 \left(z- \sqrt{z(1+z)} \right) \right] \bigg) \,.
\end{aligned}
\end{equation}

\subsection{$I_{\cal F}$ functional}

Let us consider the integral that appears in the definition \eqref{F functional} of the $I_{\cal F}$ functional, i.e.
\begin{equation}
\label{F functional-zp-1}
I_{\cal F}[z^p] \sim -\frac{1}{2\sqrt{\pi}} \int  \frac{d^D X_5}{i\pi^{D/2}} \, \left( \frac{X_{13}^2 X_{24}^2}{X_{15}^2 X_{25}^2 X_{35}^2 X_{45}^2} \right)^{\frac{3}{4}} z^p \,,
\end{equation}
with $p \in \mathbb{Z}$, $D=3-2\epsilon$, and where to simplify notation we have chosen to use the symbol $\sim$ to indicate that we are only retaining the leading $1/\epsilon^2$ divergence. Using Feynman parametrization we get
\begin{equation*}
\label{F functional-zp-2}
I_{\cal F}[z^p]  \sim -\frac{X_{13}^{3/2+2p} X_{24}^{3/2-2p}}{ \sqrt{\pi} \, \Gamma \left( \frac{3}{4}+p \right)^2 \Gamma \left( \frac{3}{4}-p \right)^2} \, \frac{1}{{\rm Vol}[{\rm GL}(1)]} \left( \prod_{i=1}^4 \int_0^{\infty} d\alpha_i \right) \, \int \frac{d^D X_5}{i\pi^{D/2}} \,  \frac{(\alpha_1 \alpha_3)^{-\frac{1}{4}+p} (\alpha_2 \alpha_4)^{-\frac{1}{4}-p} }{\left( -2X_5.W \right)^3} \,,
\end{equation*}
with 
\begin{equation}
\label{W}
W:= \sum_{i=1}^4 \alpha_i X_i  \,.
\end{equation}
Working as with the ${\cal E}$ integral discussed in the previous section we arrive at
\begin{equation*}
\label{F functional-zp-3}
I_{\cal F}[z^p]  \sim -\frac{\Gamma \left( \frac{3}{2}+\epsilon \right) \Gamma(-\epsilon)^2}{2 \sqrt{\pi} \, \Gamma \left( \frac{3}{4}+p \right)^2 \Gamma \left( \frac{3}{4}-p \right)^2 \Gamma(-2\epsilon)} \, \int_0^{1} d\gamma_2  \, \int_0^{1} d\gamma_3 \, \frac{(\gamma_2 \gamma_3)^{-\frac{1}{4}+p} [(1-\gamma_2) (1-\gamma_3)]^{-\frac{1}{4}-p} }{\left[ \gamma_2 \gamma_3 + (1-\gamma_2) (1-\gamma_3) \right]^{3/2+\epsilon}} \,.
\end{equation*}
At this point is useful to use the Mellin-Barnes formula, which allows us to write
\begin{equation}
\label{F functional-zp-4}
\begin{aligned}
I_{\cal F}[z^p]  &\sim -\frac{1}{2 \sqrt{\pi} \, \Gamma \left( \frac{3}{4}+p \right)^2 \Gamma \left( \frac{3}{4}-p \right)^2 \Gamma(-2\epsilon)} \, \times \\
& \qquad \int_{\zeta-i \infty}^{\zeta+i \infty} \frac{dz}{2\pi i} \, \Gamma \left( z+ \frac{3}{2}+\epsilon \right) \Gamma(-z) \Gamma^2 \left( \frac{3}{4} +p +z \right) \Gamma^2 \left( -\frac{3}{4} -p -z - \epsilon \right) \,,
\end{aligned}
\end{equation}
with
\begin{equation}
\label{eta constraint}
-\frac{3}{4}-p < \zeta < -\frac{3}{4}-p - \epsilon \,.
\end{equation}
To compute the leading divergence of \eqref{F functional-zp-4} we have chosen to follow the method described in \cite{Tausk:1999vh}, which was later automatized in a \textit{Mathematica} package by \cite{Czakon:2005rk}. Then, we finally get
\begin{equation}
\label{F functional-zp-final}
I_{\cal F}[z^p] = -\frac{2}{\sqrt{\pi} \, \Gamma \left( \frac{3}{4} +p \right) \Gamma \left( \frac{3}{4} -p \right)} \,.
\end{equation}

\subsection{$I_{\cal G}$ functional}

Finally, let us focus now on the integral that defines the $I_{\cal G}$ functional in \eqref{G functional}. That is, we will consider
\begin{equation}
\label{G functional-zp-1}
I_{\cal G}[z^p] \sim -\frac{1}{2 \sqrt{\pi}} \int \frac{d^D X_5}{i\pi^{D/2}} \,  \frac{i\epsilon(1,2,3,4,5)}{X_{15}^2 X_{25}^2 X_{35}^2 X_{45}^2} \, z^p \,.
\end{equation}
where again we are using the symbol $\sim$ to indicate that we are only keeping the $1/\epsilon^2$ contribution.
Introducing Feynman parameters we get
\begin{equation*}
\label{G functional-zp-2}
I_{\cal G}[z^p] \sim -\frac{i X_{13}^{2p} X_{24}^{-2p} \, \epsilon_{\mu \nu \rho \sigma \eta} X_1^{\mu} X_2^{\nu} X_3^{\rho} X_4^{\sigma}}{2 \sqrt{\pi} \, \Gamma^2(1+p) \, \Gamma^2(1-p) \, \text{Vol}[\text{GL(1)}] }  \left( \prod_{i=1}^4 \int_{0}^{\infty} d\alpha_i \, \right) \, \left( \frac{\alpha_1 \alpha_3}{\alpha_2 \alpha_4} \right)^{p}\, \partial^{\eta}_W \left[ \int \frac{d^D X_5}{i\pi^{D/2}} \,  \frac{1}{(-2 X_5.W)^3} \right] \,,
\end{equation*}
where $W$ was defined in \eqref{W}. The integral 
\begin{equation}
\label{G functional-epsilon integral}
\epsilon_{\mu \nu \rho \sigma \eta} X_1^{\mu} X_2^{\nu} X_3^{\rho} X_4^{\sigma} \, \partial^{\eta}_W \left[ \int \frac{d^D X_5}{i\pi^{D/2}} \,  \frac{1}{(-2 X_5.W)^3} \right] \,,
\end{equation}
was solved in \cite{Chen:2011vv,Caron-Huot:2012sos} using a regularization scheme that allows one to dimensionally regularize the integral without losing the projective invariance that comes from the constraint \eqref{5D light-cone-1}, getting as a result that \eqref{G functional-epsilon integral} is $\mathcal{O}(\epsilon)$. Therefore,
\begin{equation}
\label{G functional-zp-final}
I_{\cal G}[z^p] =0 \,.
\end{equation}

%%%%%%%%%%%%%%%%%%%%%%%%%%%%%%%%%%%%%%
\bibliographystyle{ssg}
\bibliography{ABJM_ref.bib}

\begingroup\raggedright\begin{thebibliography}{10}

\bibitem{Parke:1986gb}
S.~J. Parke and T.~R. Taylor, ``{An Amplitude for $n$ Gluon Scattering},'' {\em Phys. Rev. Lett.} {\bf 56} (1986) 2459.

\bibitem{Arkani-Hamed:2013jha}
N.~Arkani-Hamed and J.~Trnka, ``{The Amplituhedron},'' {\em JHEP} {\bf 10} (2014) 030, \href{https://arxiv.org/abs/1312.2007}{{\tt 1312.2007}}.

\bibitem{Arkani-Hamed:2013kca}
N.~Arkani-Hamed and J.~Trnka, ``{Into the Amplituhedron},'' {\em JHEP} {\bf 12} (2014) 182, \href{https://arxiv.org/abs/1312.7878}{{\tt 1312.7878}}.

\bibitem{Franco:2014csa}
S.~Franco, D.~Galloni, A.~Mariotti, and J.~Trnka, ``{Anatomy of the Amplituhedron},'' {\em JHEP} {\bf 03} (2015) 128, \href{https://arxiv.org/abs/1408.3410}{{\tt 1408.3410}}.

\bibitem{Arkani-Hamed:2017vfh}
N.~Arkani-Hamed, H.~Thomas, and J.~Trnka, ``{Unwinding the Amplituhedron in Binary},'' {\em JHEP} {\bf 01} (2018) 016, \href{https://arxiv.org/abs/1704.05069}{{\tt 1704.05069}}.

\bibitem{Arkani-Hamed:2018rsk}
N.~Arkani-Hamed, C.~Langer, A.~Yelleshpur~Srikant, and J.~Trnka, ``{Deep Into the Amplituhedron: Amplitude Singularities at All Loops and Legs},'' {\em Phys. Rev. Lett.} {\bf 122} (2019), no.~5 051601, \href{https://arxiv.org/abs/1810.08208}{{\tt 1810.08208}}.

\bibitem{Damgaard:2019ztj}
D.~Damgaard, L.~Ferro, T.~Lukowski, and M.~Parisi, ``{The Momentum Amplituhedron},'' {\em JHEP} {\bf 08} (2019) 042, \href{https://arxiv.org/abs/1905.04216}{{\tt 1905.04216}}.

\bibitem{Ferro:2022abq}
L.~Ferro and T.~Lukowski, ``{The Loop Momentum Amplituhedron},'' \href{https://arxiv.org/abs/2210.01127}{{\tt 2210.01127}}.

\bibitem{Arkani-Hamed:2017mur}
N.~Arkani-Hamed, Y.~Bai, S.~He, and G.~Yan, ``{Scattering Forms and the Positive Geometry of Kinematics, Color and the Worldsheet},'' {\em JHEP} {\bf 05} (2018) 096, \href{https://arxiv.org/abs/1711.09102}{{\tt 1711.09102}}.

\bibitem{Arkani-Hamed:2017fdk}
N.~Arkani-Hamed, P.~Benincasa, and A.~Postnikov, ``{Cosmological Polytopes and the Wavefunction of the Universe},'' \href{https://arxiv.org/abs/1709.02813}{{\tt 1709.02813}}.

\bibitem{Huang:2021jlh}
Y.-t. Huang, R.~Kojima, C.~Wen, and S.-Q. Zhang, ``{The orthogonal momentum amplituhedron and ABJM amplitudes},'' {\em JHEP} {\bf 01} (2022) 141, \href{https://arxiv.org/abs/2111.03037}{{\tt 2111.03037}}.

\bibitem{He:2021llb}
S.~He, C.-K. Kuo, and Y.-Q. Zhang, ``{The momentum amplituhedron of SYM and ABJM from twistor-string maps},'' {\em JHEP} {\bf 02} (2022) 148, \href{https://arxiv.org/abs/2111.02576}{{\tt 2111.02576}}.

\bibitem{He:2022cup}
S.~He, C.-K. Kuo, Z.~Li, and Y.-Q. Zhang, ``{All-Loop Four-Point Aharony-Bergman-Jafferis-Maldacena Amplitudes from Dimensional Reduction of the Amplituhedron},'' {\em Phys. Rev. Lett.} {\bf 129} (2022), no.~22 221604, \href{https://arxiv.org/abs/2204.08297}{{\tt 2204.08297}}.

\bibitem{Ferro:2020ygk}
L.~Ferro and T.~Lukowski, ``{Amplituhedra, and beyond},'' {\em J. Phys. A} {\bf 54} (2021), no.~3 033001, \href{https://arxiv.org/abs/2007.04342}{{\tt 2007.04342}}.

\bibitem{Herrmann:2022nkh}
E.~Herrmann and J.~Trnka, ``{The SAGEX review on scattering amplitudes Chapter 7: Positive geometry of scattering amplitudes},'' {\em J. Phys. A} {\bf 55} (2022), no.~44 443008, \href{https://arxiv.org/abs/2203.13018}{{\tt 2203.13018}}.

\bibitem{Arkani-Hamed:2021iya}
N.~Arkani-Hamed, J.~Henn, and J.~Trnka, ``{Nonperturbative negative geometries: amplitudes at strong coupling and the amplituhedron},'' {\em JHEP} {\bf 03} (2022) 108, \href{https://arxiv.org/abs/2112.06956}{{\tt 2112.06956}}.

\bibitem{Alday:2011ga}
L.~F. Alday, E.~I. Buchbinder, and A.~A. Tseytlin, ``{Correlation function of null polygonal Wilson loops with local operators},'' {\em JHEP} {\bf 09} (2011) 034, \href{https://arxiv.org/abs/1107.5702}{{\tt 1107.5702}}.

\bibitem{Adamo:2011cd}
T.~Adamo, ``{Correlation functions, null polygonal Wilson loops, and local operators},'' {\em JHEP} {\bf 12} (2011) 006, \href{https://arxiv.org/abs/1110.3925}{{\tt 1110.3925}}.

\bibitem{Engelund:2011fg}
O.~T. Engelund and R.~Roiban, ``{On correlation functions of Wilson loops, local and non-local operators},'' {\em JHEP} {\bf 05} (2012) 158, \href{https://arxiv.org/abs/1110.0758}{{\tt 1110.0758}}.

\bibitem{Alday:2012hy}
L.~F. Alday, P.~Heslop, and J.~Sikorowski, ``{Perturbative correlation functions of null Wilson loops and local operators},'' {\em JHEP} {\bf 03} (2013) 074, \href{https://arxiv.org/abs/1207.4316}{{\tt 1207.4316}}.

\bibitem{Alday:2013ip}
L.~F. Alday, J.~M. Henn, and J.~Sikorowski, ``{Higher loop mixed correlators in N=4 SYM},'' {\em JHEP} {\bf 03} (2013) 058, \href{https://arxiv.org/abs/1301.0149}{{\tt 1301.0149}}.

\bibitem{Hernandez:2013kb}
R.~Hernandez and J.~M. Nieto, ``{Holographic correlation functions of hexagon Wilson loops with one local operator},'' {\em Phys. Lett. B} {\bf 726} (2013) 417--421, \href{https://arxiv.org/abs/1301.7220}{{\tt 1301.7220}}.

\bibitem{Henn:2019swt}
J.~M. Henn, G.~P. Korchemsky, and B.~Mistlberger, ``{The full four-loop cusp anomalous dimension in $\mathcal{N}=4$ super Yang-Mills and QCD},'' {\em JHEP} {\bf 04} (2020) 018, \href{https://arxiv.org/abs/1911.10174}{{\tt 1911.10174}}.

\bibitem{Chicherin:2022bov}
D.~Chicherin and J.~M. Henn, ``{Symmetry properties of Wilson loops with a Lagrangian insertion},'' {\em JHEP} {\bf 07} (2022) 057, \href{https://arxiv.org/abs/2202.05596}{{\tt 2202.05596}}.

\bibitem{Chicherin:2022zxo}
D.~Chicherin and J.~Henn, ``{Pentagon Wilson loop with Lagrangian insertion at two loops in $ \mathcal{N} $ = 4 super Yang-Mills theory},'' {\em JHEP} {\bf 07} (2022) 038, \href{https://arxiv.org/abs/2204.00329}{{\tt 2204.00329}}.

\bibitem{Alday:2007hr}
L.~F. Alday and J.~M. Maldacena, ``{Gluon scattering amplitudes at strong coupling},'' {\em JHEP} {\bf 06} (2007) 064, \href{https://arxiv.org/abs/0705.0303}{{\tt 0705.0303}}.

\bibitem{Drummond:2007aua}
J.~M. Drummond, G.~P. Korchemsky, and E.~Sokatchev, ``{Conformal properties of four-gluon planar amplitudes and Wilson loops},'' {\em Nucl. Phys. B} {\bf 795} (2008) 385--408, \href{https://arxiv.org/abs/0707.0243}{{\tt 0707.0243}}.

\bibitem{Brandhuber:2007yx}
A.~Brandhuber, P.~Heslop, and G.~Travaglini, ``{MHV amplitudes in N=4 super Yang-Mills and Wilson loops},'' {\em Nucl. Phys. B} {\bf 794} (2008) 231--243, \href{https://arxiv.org/abs/0707.1153}{{\tt 0707.1153}}.

\bibitem{Drummond:2007cf}
J.~M. Drummond, J.~Henn, G.~P. Korchemsky, and E.~Sokatchev, ``{On planar gluon amplitudes/Wilson loops duality},'' {\em Nucl. Phys. B} {\bf 795} (2008) 52--68, \href{https://arxiv.org/abs/0709.2368}{{\tt 0709.2368}}.

\bibitem{Beisert:2006ez}
N.~Beisert, B.~Eden, and M.~Staudacher, ``{Transcendentality and Crossing},'' {\em J. Stat. Mech.} {\bf 0701} (2007) P01021, \href{https://arxiv.org/abs/hep-th/0610251}{{\tt hep-th/0610251}}.

\bibitem{Correa:2012hh}
D.~Correa, J.~Maldacena, and A.~Sever, ``{The quark anti-quark potential and the cusp anomalous dimension from a TBA equation},'' {\em JHEP} {\bf 08} (2012) 134, \href{https://arxiv.org/abs/1203.1913}{{\tt 1203.1913}}.

\bibitem{Aharony:2008ug}
O.~Aharony, O.~Bergman, D.~L. Jafferis, and J.~Maldacena, ``{N=6 superconformal Chern-Simons-matter theories, M2-branes and their gravity duals},'' {\em JHEP} {\bf 10} (2008) 091, \href{https://arxiv.org/abs/0806.1218}{{\tt 0806.1218}}.

\bibitem{Bargheer:2010hn}
T.~Bargheer, F.~Loebbert, and C.~Meneghelli, ``{Symmetries of Tree-level Scattering Amplitudes in N=6 Superconformal Chern-Simons Theory},'' {\em Phys. Rev. D} {\bf 82} (2010) 045016, \href{https://arxiv.org/abs/1003.6120}{{\tt 1003.6120}}.

\bibitem{Chen:2011vv}
W.-M. Chen and Y.-t. Huang, ``{Dualities for Loop Amplitudes of N=6 Chern-Simons Matter Theory},'' {\em JHEP} {\bf 11} (2011) 057, \href{https://arxiv.org/abs/1107.2710}{{\tt 1107.2710}}.

\bibitem{Bargheer:2012cp}
T.~Bargheer, N.~Beisert, F.~Loebbert, and T.~McLoughlin, ``{Conformal Anomaly for Amplitudes in $\mathcal{N}=6$ Superconformal Chern-Simons Theory},'' {\em J. Phys. A} {\bf 45} (2012) 475402, \href{https://arxiv.org/abs/1204.4406}{{\tt 1204.4406}}.

\bibitem{Leoni:2015zxa}
M.~Leoni, A.~Mauri, and A.~Santambrogio, ``{On the amplitude/Wilson loop duality in N=2 SCQCD},'' {\em Phys. Lett. B} {\bf 747} (2015) 325--330, \href{https://arxiv.org/abs/1502.07614}{{\tt 1502.07614}}.

\bibitem{Bianchi:2014iia}
M.~S. Bianchi and M.~Leoni, ``{On the ABJM four-point amplitude at three loops and BDS exponentiation},'' {\em JHEP} {\bf 11} (2014) 077, \href{https://arxiv.org/abs/1403.3398}{{\tt 1403.3398}}.

\bibitem{Bianchi:2011aa}
M.~S. Bianchi, M.~Leoni, and S.~Penati, ``{An All Order Identity between ABJM and N=4 SYM Four-Point Amplitudes},'' {\em JHEP} {\bf 04} (2012) 045, \href{https://arxiv.org/abs/1112.3649}{{\tt 1112.3649}}.

\bibitem{Bianchi:2013iha}
M.~S. Bianchi, M.~Leoni, M.~Leoni, A.~Mauri, S.~Penati, and A.~Santambrogio, ``{ABJM amplitudes and WL at finite $N$},'' {\em JHEP} {\bf 09} (2013) 114, \href{https://arxiv.org/abs/1306.3243}{{\tt 1306.3243}}.

\bibitem{Bianchi:2013pfa}
L.~Bianchi and M.~S. Bianchi, ``{Nonplanarity through unitarity in the ABJM theory},'' {\em Phys. Rev. D} {\bf 89} (2014), no.~12 125002, \href{https://arxiv.org/abs/1311.6464}{{\tt 1311.6464}}.

\bibitem{Caron-Huot:2012sos}
S.~Caron-Huot and Y.-t. Huang, ``{The two-loop six-point amplitude in ABJM theory},'' {\em JHEP} {\bf 03} (2013) 075, \href{https://arxiv.org/abs/1210.4226}{{\tt 1210.4226}}.

\bibitem{He:2022lfz}
S.~He, Y.-t. Huang, C.-K. Kuo, and Z.~Li, ``{The two-loop eight-point amplitude in ABJM theory},'' \href{https://arxiv.org/abs/2211.01792}{{\tt 2211.01792}}.

\bibitem{Bianchi:2012cq}
M.~S. Bianchi, M.~Leoni, A.~Mauri, S.~Penati, and A.~Santambrogio, ``{One Loop Amplitudes In ABJM},'' {\em JHEP} {\bf 07} (2012) 029, \href{https://arxiv.org/abs/1204.4407}{{\tt 1204.4407}}.

\bibitem{Henn:2010ps}
J.~M. Henn, J.~Plefka, and K.~Wiegandt, ``{Light-like polygonal Wilson loops in 3d Chern-Simons and ABJM theory},'' {\em JHEP} {\bf 08} (2010) 032, \href{https://arxiv.org/abs/1004.0226}{{\tt 1004.0226}}. [Erratum: JHEP 11, 053 (2011)].

\bibitem{Bianchi:2011dg}
M.~S. Bianchi, M.~Leoni, A.~Mauri, S.~Penati, and A.~Santambrogio, ``{Scattering Amplitudes/Wilson Loop Duality In ABJM Theory},'' {\em JHEP} {\bf 01} (2012) 056, \href{https://arxiv.org/abs/1107.3139}{{\tt 1107.3139}}.

\bibitem{Huang:2010qy}
Y.-t. Huang and A.~E. Lipstein, ``{Dual Superconformal Symmetry of N=6 Chern-Simons Theory},'' {\em JHEP} {\bf 11} (2010) 076, \href{https://arxiv.org/abs/1008.0041}{{\tt 1008.0041}}.

\bibitem{Gang:2010gy}
D.~Gang, Y.-t. Huang, E.~Koh, S.~Lee, and A.~E. Lipstein, ``{Tree-level Recursion Relation and Dual Superconformal Symmetry of the ABJM Theory},'' {\em JHEP} {\bf 03} (2011) 116, \href{https://arxiv.org/abs/1012.5032}{{\tt 1012.5032}}.

\bibitem{Bianchi:2011fc}
M.~S. Bianchi, M.~Leoni, A.~Mauri, S.~Penati, and A.~Santambrogio, ``{Scattering in ABJ theories},'' {\em JHEP} {\bf 12} (2011) 073, \href{https://arxiv.org/abs/1110.0738}{{\tt 1110.0738}}.

\bibitem{Lee:2010du}
S.~Lee, ``{Yangian Invariant Scattering Amplitudes in Supersymmetric Chern-Simons Theory},'' {\em Phys. Rev. Lett.} {\bf 105} (2010) 151603, \href{https://arxiv.org/abs/1007.4772}{{\tt 1007.4772}}.

\bibitem{Gromov:2008qe}
N.~Gromov and P.~Vieira, ``{The all loop AdS4/CFT3 Bethe ansatz},'' {\em JHEP} {\bf 01} (2009) 016, \href{https://arxiv.org/abs/0807.0777}{{\tt 0807.0777}}.

\bibitem{Griguolo:2012iq}
L.~Griguolo, D.~Marmiroli, G.~Martelloni, and D.~Seminara, ``{The generalized cusp in ABJ(M) N = 6 Super Chern-Simons theories},'' {\em JHEP} {\bf 05} (2013) 113, \href{https://arxiv.org/abs/1208.5766}{{\tt 1208.5766}}.

\bibitem{Bianchi:2014ada}
L.~Bianchi, M.~S. Bianchi, A.~Bres, V.~Forini, and E.~Vescovi, ``{Two-loop cusp anomaly in ABJM at strong coupling},'' {\em JHEP} {\bf 10} (2014) 013, \href{https://arxiv.org/abs/1407.4788}{{\tt 1407.4788}}.

\bibitem{Beisert:2006qh}
N.~Beisert, ``{The Analytic Bethe Ansatz for a Chain with Centrally Extended su(2|2) Symmetry},'' {\em J. Stat. Mech.} {\bf 0701} (2007) P01017, \href{https://arxiv.org/abs/nlin/0610017}{{\tt nlin/0610017}}.

\bibitem{Minahan:2008hf}
J.~A. Minahan and K.~Zarembo, ``{The Bethe ansatz for superconformal Chern-Simons},'' {\em JHEP} {\bf 09} (2008) 040, \href{https://arxiv.org/abs/0806.3951}{{\tt 0806.3951}}.

\bibitem{Minahan:2009te}
J.~A. Minahan, W.~Schulgin, and K.~Zarembo, ``{Two loop integrability for Chern-Simons theories with N=6 supersymmetry},'' {\em JHEP} {\bf 03} (2009) 057, \href{https://arxiv.org/abs/0901.1142}{{\tt 0901.1142}}.

\bibitem{Bak:2008vd}
D.~Bak, D.~Gang, and S.-J. Rey, ``{Integrable Spin Chain of Superconformal U(M) x anti-U(N) Chern-Simons Theory},'' {\em JHEP} {\bf 10} (2008) 038, \href{https://arxiv.org/abs/0808.0170}{{\tt 0808.0170}}.

\bibitem{Minahan:2009aq}
J.~A. Minahan, O.~Ohlsson~Sax, and C.~Sieg, ``{Magnon dispersion to four loops in the ABJM and ABJ models},'' {\em J. Phys. A} {\bf 43} (2010) 275402, \href{https://arxiv.org/abs/0908.2463}{{\tt 0908.2463}}.

\bibitem{Minahan:2009wg}
J.~A. Minahan, O.~Ohlsson~Sax, and C.~Sieg, ``{Anomalous dimensions at four loops in N=6 superconformal Chern-Simons theories},'' {\em Nucl. Phys. B} {\bf 846} (2011) 542--606, \href{https://arxiv.org/abs/0912.3460}{{\tt 0912.3460}}.

\bibitem{Leoni:2010tb}
M.~Leoni, A.~Mauri, J.~A. Minahan, O.~Ohlsson~Sax, A.~Santambrogio, C.~Sieg, and G.~Tartaglino-Mazzucchelli, ``{Superspace calculation of the four-loop spectrum in N=6 supersymmetric Chern-Simons theories},'' {\em JHEP} {\bf 12} (2010) 074, \href{https://arxiv.org/abs/1010.1756}{{\tt 1010.1756}}.

\bibitem{Gromov:2014eha}
N.~Gromov and G.~Sizov, ``{Exact Slope and Interpolating Functions in N=6 Supersymmetric Chern-Simons Theory},'' {\em Phys. Rev. Lett.} {\bf 113} (2014), no.~12 121601, \href{https://arxiv.org/abs/1403.1894}{{\tt 1403.1894}}.

\bibitem{Cavaglia:2016ide}
A.~Cavagli\`a, N.~Gromov, and F.~Levkovich-Maslyuk, ``{On the Exact Interpolating Function in ABJ Theory},'' {\em JHEP} {\bf 12} (2016) 086, \href{https://arxiv.org/abs/1605.04888}{{\tt 1605.04888}}.

\bibitem{Hodges:2009hk}
A.~Hodges, ``{Eliminating spurious poles from gauge-theoretic amplitudes},'' {\em JHEP} {\bf 05} (2013) 135, \href{https://arxiv.org/abs/0905.1473}{{\tt 0905.1473}}.

\bibitem{Bern:2012uc}
Z.~Bern, J.~J.~M. Carrasco, H.~Johansson, and R.~Roiban, ``{The Five-Loop Four-Point Amplitude of N=4 super-Yang-Mills Theory},'' {\em Phys. Rev. Lett.} {\bf 109} (2012) 241602, \href{https://arxiv.org/abs/1207.6666}{{\tt 1207.6666}}.

\bibitem{Arkani-Hamed:2010zjl}
N.~Arkani-Hamed, J.~L. Bourjaily, F.~Cachazo, S.~Caron-Huot, and J.~Trnka, ``{The All-Loop Integrand For Scattering Amplitudes in Planar N=4 SYM},'' {\em JHEP} {\bf 01} (2011) 041, \href{https://arxiv.org/abs/1008.2958}{{\tt 1008.2958}}.

\bibitem{Bourjaily:2011hi}
J.~L. Bourjaily, A.~DiRe, A.~Shaikh, M.~Spradlin, and A.~Volovich, ``{The Soft-Collinear Bootstrap: N=4 Yang-Mills Amplitudes at Six and Seven Loops},'' {\em JHEP} {\bf 03} (2012) 032, \href{https://arxiv.org/abs/1112.6432}{{\tt 1112.6432}}.

\bibitem{Bourjaily:2016evz}
J.~L. Bourjaily, P.~Heslop, and V.-V. Tran, ``{Amplitudes and Correlators to Ten Loops Using Simple, Graphical Bootstraps},'' {\em JHEP} {\bf 11} (2016) 125, \href{https://arxiv.org/abs/1609.00007}{{\tt 1609.00007}}.

\bibitem{Eden:2012tu}
B.~Eden, P.~Heslop, G.~P. Korchemsky, and E.~Sokatchev, ``{Constructing the correlation function of four stress-tensor multiplets and the four-particle amplitude in N=4 SYM},'' {\em Nucl. Phys. B} {\bf 862} (2012) 450--503, \href{https://arxiv.org/abs/1201.5329}{{\tt 1201.5329}}.

\bibitem{Arkani-Hamed:2010pyv}
N.~Arkani-Hamed, J.~L. Bourjaily, F.~Cachazo, and J.~Trnka, ``{Local Integrals for Planar Scattering Amplitudes},'' {\em JHEP} {\bf 06} (2012) 125, \href{https://arxiv.org/abs/1012.6032}{{\tt 1012.6032}}.

\bibitem{Agarwal:2008pu}
A.~Agarwal, N.~Beisert, and T.~McLoughlin, ``{Scattering in Mass-Deformed N\ensuremath{>}=4 Chern-Simons Models},'' {\em JHEP} {\bf 06} (2009) 045, \href{https://arxiv.org/abs/0812.3367}{{\tt 0812.3367}}.

\bibitem{Korchemskaya:1992je}
I.~A. Korchemskaya and G.~P. Korchemsky, ``{On lightlike Wilson loops},'' {\em Phys. Lett. B} {\bf 287} (1992) 169--175.

\bibitem{Gromov:2012eu}
N.~Gromov and A.~Sever, ``{Analytic Solution of Bremsstrahlung TBA},'' {\em JHEP} {\bf 11} (2012) 075, \href{https://arxiv.org/abs/1207.5489}{{\tt 1207.5489}}.

\bibitem{Eden:2010zz}
B.~Eden, G.~P. Korchemsky, and E.~Sokatchev, ``{From correlation functions to scattering amplitudes},'' {\em JHEP} {\bf 12} (2011) 002, \href{https://arxiv.org/abs/1007.3246}{{\tt 1007.3246}}.

\bibitem{Henn:2013pwa}
J.~M. Henn, ``{Multiloop integrals in dimensional regularization made simple},'' {\em Phys. Rev. Lett.} {\bf 110} (2013) 251601, \href{https://arxiv.org/abs/1304.1806}{{\tt 1304.1806}}.

\bibitem{Henn:2020omi}
J.~M. Henn, ``{What Can We Learn About QCD and Collider Physics from N=4 Super Yang\textendash{}Mills?},'' {\em Ann. Rev. Nucl. Part. Sci.} {\bf 71} (2021) 87--112, \href{https://arxiv.org/abs/2006.00361}{{\tt 2006.00361}}.

\bibitem{Bern:2005iz}
Z.~Bern, L.~J. Dixon, and V.~A. Smirnov, ``{Iteration of planar amplitudes in maximally supersymmetric Yang-Mills theory at three loops and beyond},'' {\em Phys. Rev. D} {\bf 72} (2005) 085001, \href{https://arxiv.org/abs/hep-th/0505205}{{\tt hep-th/0505205}}.

\bibitem{Dixon:2011pw}
L.~J. Dixon, J.~M. Drummond, and J.~M. Henn, ``{Bootstrapping the three-loop hexagon},'' {\em JHEP} {\bf 11} (2011) 023, \href{https://arxiv.org/abs/1108.4461}{{\tt 1108.4461}}.

\bibitem{Arkani-Hamed:2012zlh}
N.~Arkani-Hamed, J.~L. Bourjaily, F.~Cachazo, A.~B. Goncharov, A.~Postnikov, and J.~Trnka, {\em {Grassmannian Geometry of Scattering Amplitudes}}.
\newblock Cambridge University Press, 4, 2016.

\bibitem{Kotikov:2006ts}
A.~V. Kotikov and L.~N. Lipatov, ``{On the highest transcendentality in N=4 SUSY},'' {\em Nucl. Phys. B} {\bf 769} (2007) 217--255, \href{https://arxiv.org/abs/hep-th/0611204}{{\tt hep-th/0611204}}.

\bibitem{Kotikov:2002ab}
A.~V. Kotikov and L.~N. Lipatov, ``{DGLAP and BFKL equations in the $N=4$ supersymmetric gauge theory},'' {\em Nucl. Phys. B} {\bf 661} (2003) 19--61, \href{https://arxiv.org/abs/hep-ph/0208220}{{\tt hep-ph/0208220}}. [Erratum: Nucl.Phys.B 685, 405--407 (2004)].

\bibitem{Kotikov:2004er}
A.~V. Kotikov, L.~N. Lipatov, A.~I. Onishchenko, and V.~N. Velizhanin, ``{Three loop universal anomalous dimension of the Wilson operators in $N=4$ SUSY Yang-Mills model},'' {\em Phys. Lett. B} {\bf 595} (2004) 521--529, \href{https://arxiv.org/abs/hep-th/0404092}{{\tt hep-th/0404092}}. [Erratum: Phys.Lett.B 632, 754--756 (2006)].

\bibitem{Bianchi:2013pva}
M.~S. Bianchi, G.~Giribet, M.~Leoni, and S.~Penati, ``{Light-like Wilson loops in ABJM and maximal transcendentality},'' {\em JHEP} {\bf 08} (2013) 111, \href{https://arxiv.org/abs/1304.6085}{{\tt 1304.6085}}.

\bibitem{Hannesdottir:2021kpd}
H.~S. Hannesdottir, A.~J. McLeod, M.~D. Schwartz, and C.~Vergu, ``{Implications of the Landau equations for iterated integrals},'' {\em Phys. Rev. D} {\bf 105} (2022), no.~6 L061701, \href{https://arxiv.org/abs/2109.09744}{{\tt 2109.09744}}.

\bibitem{Goncharov:2010jf}
A.~B. Goncharov, M.~Spradlin, C.~Vergu, and A.~Volovich, ``{Classical Polylogarithms for Amplitudes and Wilson Loops},'' {\em Phys. Rev. Lett.} {\bf 105} (2010) 151605, \href{https://arxiv.org/abs/1006.5703}{{\tt 1006.5703}}.

\bibitem{Dixon:2011nj}
L.~J. Dixon, J.~M. Drummond, and J.~M. Henn, ``{Analytic result for the two-loop six-point NMHV amplitude in N=4 super Yang-Mills theory},'' {\em JHEP} {\bf 01} (2012) 024, \href{https://arxiv.org/abs/1111.1704}{{\tt 1111.1704}}.

\bibitem{Dixon:2014voa}
L.~J. Dixon, J.~M. Drummond, C.~Duhr, and J.~Pennington, ``{The four-loop remainder function and multi-Regge behavior at NNLLA in planar N = 4 super-Yang-Mills theory},'' {\em JHEP} {\bf 06} (2014) 116, \href{https://arxiv.org/abs/1402.3300}{{\tt 1402.3300}}.

\bibitem{Dixon:2015iva}
L.~J. Dixon, M.~von Hippel, and A.~J. McLeod, ``{The four-loop six-gluon NMHV ratio function},'' {\em JHEP} {\bf 01} (2016) 053, \href{https://arxiv.org/abs/1509.08127}{{\tt 1509.08127}}.

\bibitem{Caron-Huot:2016owq}
S.~Caron-Huot, L.~J. Dixon, A.~McLeod, and M.~von Hippel, ``{Bootstrapping a Five-Loop Amplitude Using Steinmann Relations},'' {\em Phys. Rev. Lett.} {\bf 117} (2016), no.~24 241601, \href{https://arxiv.org/abs/1609.00669}{{\tt 1609.00669}}.

\bibitem{Drummond:2014ffa}
J.~M. Drummond, G.~Papathanasiou, and M.~Spradlin, ``{A Symbol of Uniqueness: The Cluster Bootstrap for the 3-Loop MHV Heptagon},'' {\em JHEP} {\bf 03} (2015) 072, \href{https://arxiv.org/abs/1412.3763}{{\tt 1412.3763}}.

\bibitem{Dixon:2016nkn}
L.~J. Dixon, J.~Drummond, T.~Harrington, A.~J. McLeod, G.~Papathanasiou, and M.~Spradlin, ``{Heptagons from the Steinmann Cluster Bootstrap},'' {\em JHEP} {\bf 02} (2017) 137, \href{https://arxiv.org/abs/1612.08976}{{\tt 1612.08976}}.

\bibitem{Huang:2013owa}
Y.-T. Huang and C.~Wen, ``{ABJM amplitudes and the positive orthogonal grassmannian},'' {\em JHEP} {\bf 02} (2014) 104, \href{https://arxiv.org/abs/1309.3252}{{\tt 1309.3252}}.

\bibitem{Huang:2014xza}
Y.-t. Huang, C.~Wen, and D.~Xie, ``{The Positive orthogonal Grassmannian and loop amplitudes of ABJM},'' {\em J. Phys. A} {\bf 47} (2014), no.~47 474008, \href{https://arxiv.org/abs/1402.1479}{{\tt 1402.1479}}.

\bibitem{He:2023exb}
S.~He, C.-K. Kuo, Z.~Li, and Y.-Q. Zhang, ``{Emergent unitarity, all-loop cuts and integrations from the ABJM amplituhedron},'' \href{https://arxiv.org/abs/2303.03035}{{\tt 2303.03035}}.

\bibitem{Tausk:1999vh}
J.~B. Tausk, ``{Nonplanar massless two loop Feynman diagrams with four on-shell legs},'' {\em Phys. Lett. B} {\bf 469} (1999) 225--234, \href{https://arxiv.org/abs/hep-ph/9909506}{{\tt hep-ph/9909506}}.

\bibitem{Czakon:2005rk}
M.~Czakon, ``{Automatized analytic continuation of Mellin-Barnes integrals},'' {\em Comput. Phys. Commun.} {\bf 175} (2006) 559--571, \href{https://arxiv.org/abs/hep-ph/0511200}{{\tt hep-ph/0511200}}.

\end{thebibliography}\endgroup
%%%%%%%%%%%%%%%%%%%%%%%%%%%%%%%%%%%%%%

\end{document}